# Structural Relaxation Enables Millisecond Infrared Photodetection in Selenium–Iodine Semiconductors

Biman Jana[1]*

[1]Indian Institute of Science, Bangalore, 560012, India

*Corresponding author, *Email:* biman.jana013@gmail.com, bimanjana@iisc.ac.in

## 1. Abstract:

The optoelectronic performance of amorphous and partially crystalline semiconductors is strongly governed by structural disorder, yet establishing direct correlations between structural evolution and carrier transport remains challenging. Here, we show that the solid-state optoelectronic response of selenium–iodine (Se-$I_2$) is dictated by its transformation from a metastable glassy phase into an ordered lamellar structure. Vertically resolved transport measurements reveal pronounced depth-dependent electrical conductivity within thick Se-$I_2$ films, arising from compositional and structural inhomogeneity developed during solidification. Immediately after solidification, the glassy Se-$I_2$ network exhibits sluggish carrier transport and persistent photoconductivity. Upon prolonged structural relaxation, however, the material undergoes significant transport enhancement, enabling millisecond-scale infrared photodetection with a response time of 4.3 ms, stable operation up to 8 kHz modulation frequency, and shot-noise-limited detectivities of ~$10^8$ Jones at 1550 nm and ~$10^{11}$ Jones in the visible. Spatial photocurrent mapping demonstrates highly uniform carrier collection despite the self-assembled lamellar morphology, while infrared thermal imaging confirms practical imaging capability over a broad temperature range. These results establish structural relaxation as a powerful strategy for engineering carrier transport in selenium-based semiconductors and position Se-$I_2$ as a promising solution-processable platform for infrared photodetection and thermal imaging.

# 2. Introduction:

Infrared photodetectors have been studied for decades due to their wide-ranging applications in telecommunications[1–3], imaging [4–6], sensing, medical diagnostics[7], environmental and safety monitoring[8], defense systems, and emerging quantum technologies[9]. Se-$I_2$, a low melting point semiconductor[10], has primarily been explored in its liquid state, and its optoelectronic response in solid liquid mixture has previously been demonstrated mainly in the visible region. In accordance with previous studies[11,12] we find that in solid state the material has structurally separate crystallites of selenium and iodine. Hence, we denote it as Se-$I_2$ instead of $SeI_2$. In its liquid state, Se-$I_2$ is observed to be flowing while preserving the semiconducting properties[13]. This previous study has also shown that upon cooling, liquid $SeI_2$ forms a glassy phase[13] that gradually recovers its crystalline character over time. This gradual recovery of crystallinity is also reflected in increase in dark current over time (see Figure S1-S3 and discussion in supplementary section 1). This suggests the possibility of improved optoelectronic performance in the solid state as structural ordering progresses over time; however, this potential enhancement of optoelectronics has not yet been systematically investigated. In addition, the influence of device geometry on charge-carrier extraction remains insufficiently understood. In this work, we observe slow bolometric photo response (~20 s) characteristics in vertical device architectures and identify a pronounced variation in doping concentration across the vertical thickness of a typical Se-$I_2$ film. By accounting for this vertical carrier-concentration gradient and taking advantage of the slow crystallization process that improves structural ordering over time, we achieve millisecond-scale temporal response in a horizontal device geometry. Our results indicate that optimizing device geometry, enhancing crystal ordering, and controlling doping concentration are critical to achieving high-performance infrared photodetectors based on this material system.

# 3. Result and Discussion:

Structural characterization (Supplementary Section 2) confirms that, unlike the molten eutectic where selenium–iodine compounds have been reported, solid-state Se-$I_2$ consists of distinct selenium and iodine phases, consistent with previous studies. [11,12] Establishing this structural distinction is important because the optoelectronic response investigated here originates from the solid-phase material rather than from selenium–iodine compounds present in the liquid state.

## 3.2 Vertical Device Response:

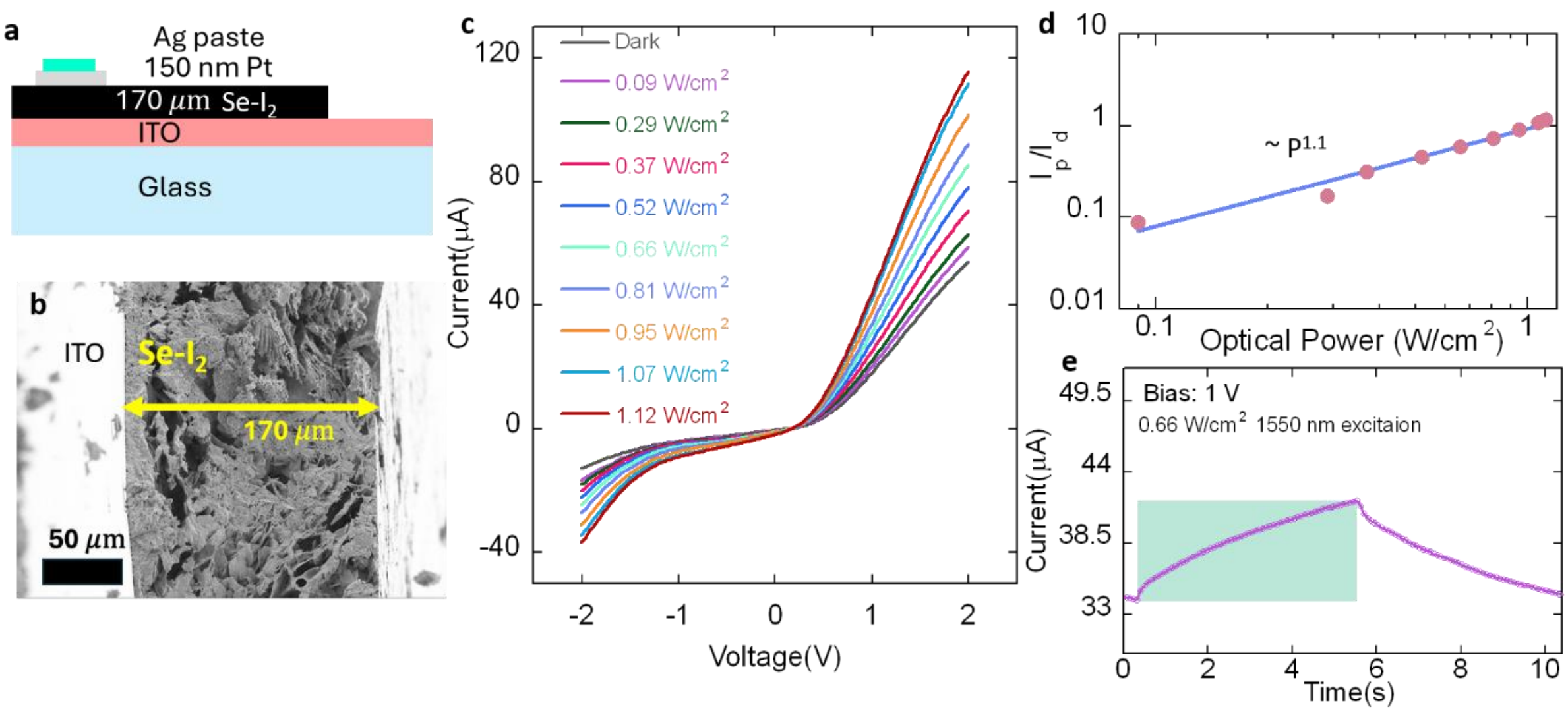

**Figure 1:** Vertical device photoresponse: **(a)** Schematic of vertical device **(b)** Cross sectional SEM image of the vertical device to measure the thickness of the $Se_nI_2$ sample on ITO substrate. **(c)** I-V characteristics both in dark and illumination with different powers of 1550 nm source. **(d)** Power dependence of $I_p/I_d$ with varying incident optical power of 1550 nm source **(e)** Response to light is shown as a function of time displaying a very slow bolometric rise and decay. The optical shutter was chopped with a frequency of 0.2 Hz. The light blue shaded region represents light exposure with 0.66 W/cm$^2$ of 1550 nm source with a vertically applied bias of 1V.

To investigate photoconductivity in vertical geometry, we fabricated the device architecture illustrated in Figure 1a (see Supplementary Section 3 for fabrication details). Forming a stable electrical contact to Se-$I_2$ is challenging because the material possesses a rough, mechanically fragile surface and readily reacts with several commonly used electrode metals. Platinum was therefore selected as the contact material owing to its chemical inertness toward Se-$I_2$ under ambient conditions, whereas Au, Ag, and Al were found to react with the material during contact formation in presence of atmospheric moisture. However, a sputtered Pt layer with a thickness of 150 nm did not form a sufficiently continuous probing electrode because of the surface roughness of Se-$I_2$ (Supplementary Section 4 and Figure S7). To ensure reliable electrical contact, a conductive silver paste was subsequently applied on top of the Pt layer. Direct deposition of silver paste onto Se-$I_2$ film was also avoided because residual free iodine rapidly reacts with Ag to form AgI [14], leading to poor electrical contact and negligible photoresponse (Figure S7a). The Pt/Ag bilayer therefore provides a chemically stable and electrically reliable contact for investigating the intrinsic optoelectronic properties of solid-state Se-$I_2$.

The vertical photodetector was fabricated on an ITO-coated glass substrate, with a Se-$I_2$ active layer of approximately 170 μm thickness, as determined from cross-sectional scanning electron microscopy (Figure 1b). The chemically stable Pt/Ag top contact described above completed the ITO/ Se-$I_2$/Pt/Ag device architecture. Illumination was introduced through the transparent ITO electrode to ensure efficient optical coupling into the absorber layer.

The current–voltage characteristics of the device under dark and 1550 nm illumination are presented in Figure 1c. In both cases, the device exhibits nonlinear current–voltage behavior, indicative of carrier transport governed by multiple interfacial barriers rather than purely ohmic conduction. Such behavior is expected in the present vertical heterostructure, where carrier injection and extraction occur across several material interfaces, including the ITO/ Se-$I_2$ and Se-$I_2$/Pt junctions. [15–17] Upon illumination, the current increases systematically with increasing optical power, confirming efficient photocarrier generation within the Se-$I_2$ absorber.

The photocurrent exhibits an approximately linear dependence on incident optical power, following $I_p/I_d \propto P^{1.1}$ (Figure 1d), where (where $I_p$ and $I_d$ denote the photocurrent and dark current, respectively. An exponent close to unity suggests that photocarrier generation scales nearly proportionally with optical excitation over the investigated power range, with only a minor contribution from trap-assisted recombination[18]. To examine whether the observed behavior is influenced by the bottom contact, an additional device incorporating a $TiO_2$ electron-transport layer between the ITO electrode and the Se-$I_2$ absorber was fabricated (Figure S8a). This device displayed similar electrical and optoelectronic characteristics, indicating that the measured photoresponse is predominantly governed by the intrinsic properties of the Se-$I_2$ layer rather than by the specific choice of bottom contact. The extracted device parameters are plotted in Figure S8b–d**,** and their definitions are provided in Supplementary Section 5**.**

Despite the robust photoresponse, both vertical device architectures exhibit relatively slow temporal response (Figure 1e). Considering the unusually large thickness of the Se-$I_2$ layer (~170 μm), we hypothesize that this behaviour originates from compositional or doping inhomogeneities across the film thickness, which can give rise to spatially varying carrier transport and trapping characteristics. To investigate this possibility, we next probe the existence of vertical doping gradients within the Se-$I_2$ absorber.

## 3.3 Vertical Inhomogeneity:

The sluggish temporal response observed in the vertical photodetector suggests that carrier transport through the thick Se-$I_2$ layer is spatially nonuniform. To probe the electrical homogeneity of the film along its thickness, a vertically resolved transport structure was fabricated in which two laterally separated platinum electrode pairs were positioned at different depths within the Se-$I_2$ layer, with a vertical separation of approximately 100 nm (Figure 2a). An intermediate $SiO_2$ spacer electrically isolated the two electrode levels, enabling independent measurements of the lateral conductivity at different depths within the same film.

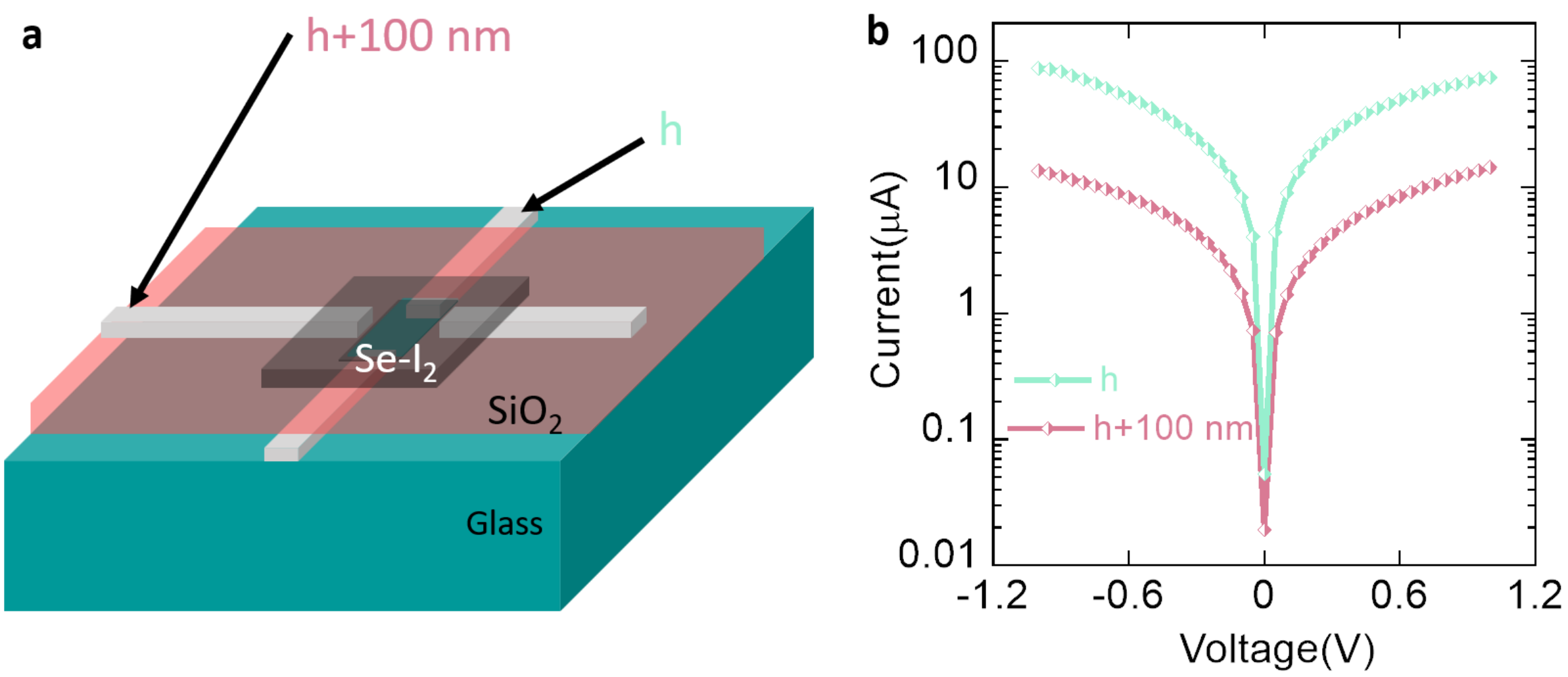

**Figure 2.** Thickness dependent electronics probe: **(a)** Device geometry is made to probe current at two different heights of Se-$I_2$ layer. Insulating $SiO_2$ layer separates the two pairs of Pt electrodes by 100 nm by height. **(b)** The current passing across the two Se-$I_2$ layers at two different heights- one at the bottom of the device and another layer 100 nm above the bottom layer.

Current–voltage measurements performed under an applied bias of 1 V reveal a pronounced conductivity gradient across the film thickness (Figure 2b). The lower electrode pair consistently exhibits nearly an order-of-magnitude higher current than the upper electrode pair, indicating substantially higher electrical conductivity in the region adjacent to the substrate. This observation demonstrates that charge transport in the thick Se-$I_2$ layer is strongly depth dependent rather than spatially uniform.

Vertical transport inhomogeneity has previously been reported in semiconductor systems containing nonuniform dopant or defect distributions, including $In_2O_3$[19], III-V nanowire[20], $Bi_2Se_3$ thin films [21], and polycrystalline CdTe [22]. Similar mechanisms are likely responsible for the behaviour observed here. During solidification, selenium and iodine separate into distinct crystalline phases (Supplementary Section 2), while molecular iodine remains volatile and readily sublimates under ambient conditions.[23] Consequently, iodine incorporation is expected to become progressively less efficient away from the substrate, producing a vertical compositional gradient across the film thickness. Such variations in iodine concentration are expected to modify both the local carrier concentration and trap density[24,25], giving rise to the observed depth-dependent electrical conductivity.

The presence of this conductivity gradient provides a plausible origin for the slow response of the vertical device. Photogenerated carriers generated in regions with lower conductivity require longer transport times before reaching the collecting electrodes, resulting in broadened transient response and incomplete carrier extraction. These observations therefore identify vertical compositional inhomogeneity as an important transport limitation in thick Se-$I_2$ photodetectors.

## 3.4 Slow Horizontal Device:

To determine whether the transport limitation originates primarily from the vertical device geometry or from the intrinsic properties of Se-$I_2$ itself, photoconductivity was next investigated using a conventional lateral device architecture. The device was fabricated by drop-casting molten Se-$I_2$ onto platinum interdigitated electrodes covering an active area of approximately 5 mm × 7 mm (Figure 3a, inset). In this configuration, photogenerated carriers are transported laterally between neighbouring electrode fingers under an applied electric field.

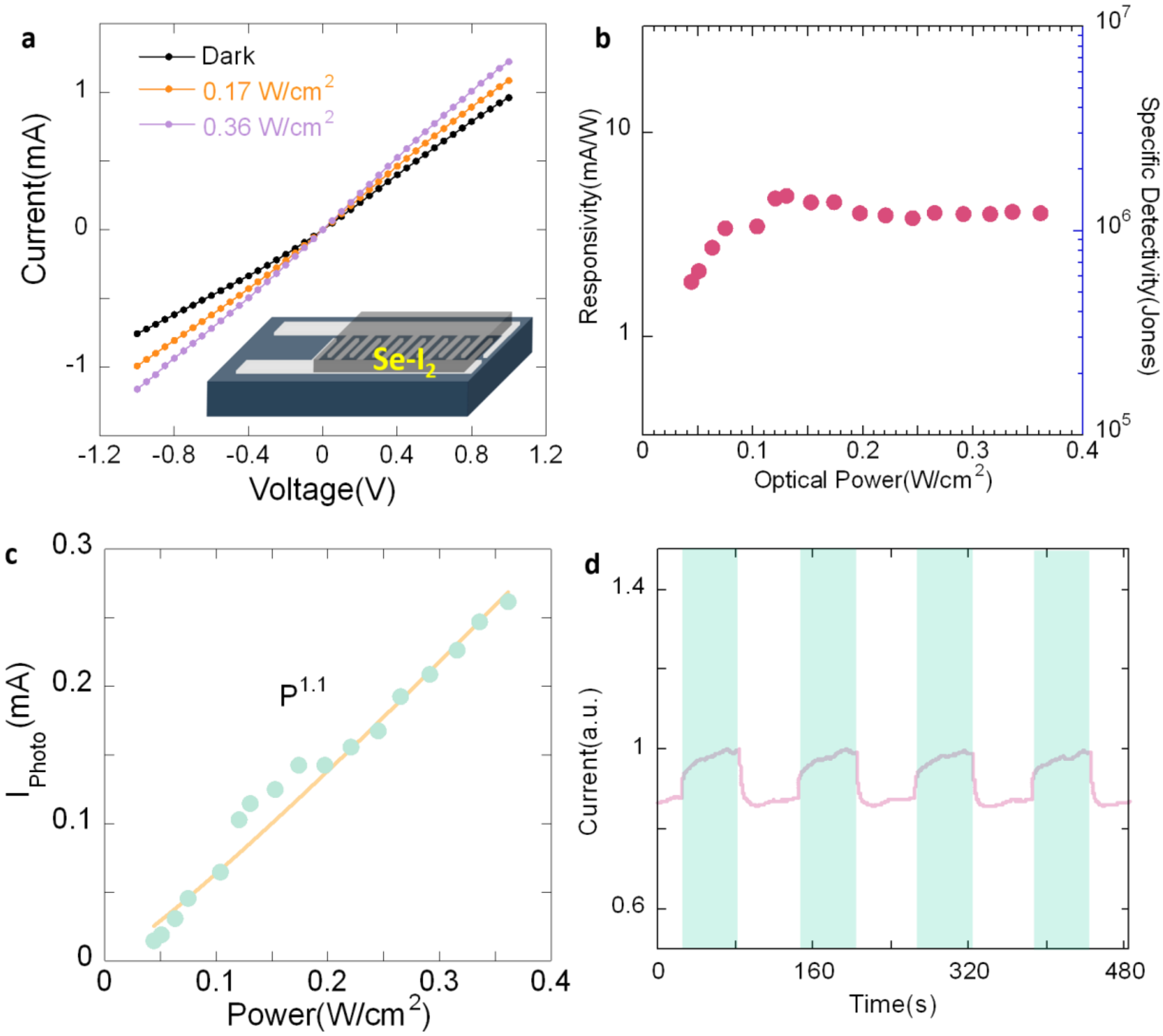

**Figure 3 (a)** Responsivity of the device with varying powers of 1550 nm. **(b)** Variation of photocurrent with incident optical power. **(c)** I-V characteristics of a large area (5 mm × 7 mm) horizontal device made by drop casting the liquid Se-$I_2$ in dark and different powers of illumination with 1550 nm source **(d)** Typical slow temporal response of a device with illumination showing a slow (~6.1 s/8.4 s) rise and decay. **(e)** Trigonal selenium unit cell with two iodine atom introduced at for chain containing four selenium atoms. **(f)** Density of states calculated for the same system shown in **(e)**.

The device exhibits a clear increase in photocurrent with increasing incident optical power (Figure 3a,b). The photocurrent follows an approximately linear power-law dependence, ($I_p \propto P^{1.1}$) (Figure 3c), indicating efficient photocarrier generation over the investigated excitation range. Despite this reproducible photoresponse, the temporal characteristics remain remarkably slow (Figure 3d and Figure S9), with response times substantially longer than those required for practical infrared photodetection.

The slow response cannot be attributed solely to the intrinsic semiconductor properties of selenium. Although the relatively large device area increases the junction capacitance and RC time constant, these electrical factors alone are insufficient to explain the pronounced response delay. Instead, the dominant limitation appears to arise from inefficient carrier transport within the Se-$I_2$ layer itself. Immediately after solidification from the melt, Se-$I_2$ adopts a metastable glassy structure, as demonstrated by the thermal and structural characterization presented earlier.[13] The resulting structural disorder introduces a broad distribution of localized electronic states [26,27] that trap photogenerated carriers and substantially prolong carrier release, thereby producing persistent photoconductivity[28,29] and sluggish temporal response.

To verify whether structural disorder is responsible for the slow carrier dynamics, the devices were subsequently subjected to prolonged aging under reduced pressure (~1 mbar). This treatment promotes gradual structural relaxation and crystallization of the initially glassy Se-$I_2$ phase. Depending on the individual sample, aging periods ranging from several months to approximately two years were required to recover long-range structural order. As shown in the following section, this structural evolution fundamentally transfoms the transport dynamics of Se-$I_2$, enabling a transition from persistent photoconductivity to millisecond-scale infrared photodetection.

## 3.5 Characterisation of the Horizontal Faster Device:

To determine how structural ordering influences carrier transport, photodetectors fabricated from aged Se-$I_2$ films were investigated after prolonged structural relaxation under reduced pressure. As discussed in the previous section, aging promotes the gradual transformation of the initially glassy Se–$I_2$ network into a more ordered crystalline state, thereby reducing structural disorder and improving electronic transport.

The optimized lateral device architecture is illustrated in Figure 4a. A Se-$I_2$ layer with a thickness of approximately 17 μm was deposited onto platinum interdigitated electrodes (10 µm finger width and spacing) and illuminated through the transparent optical path after reflection from a silver mirror. The reduced film thickness and well-defined lateral transport geometry minimize carrier transit distance while providing a uniform electric field across the active region.

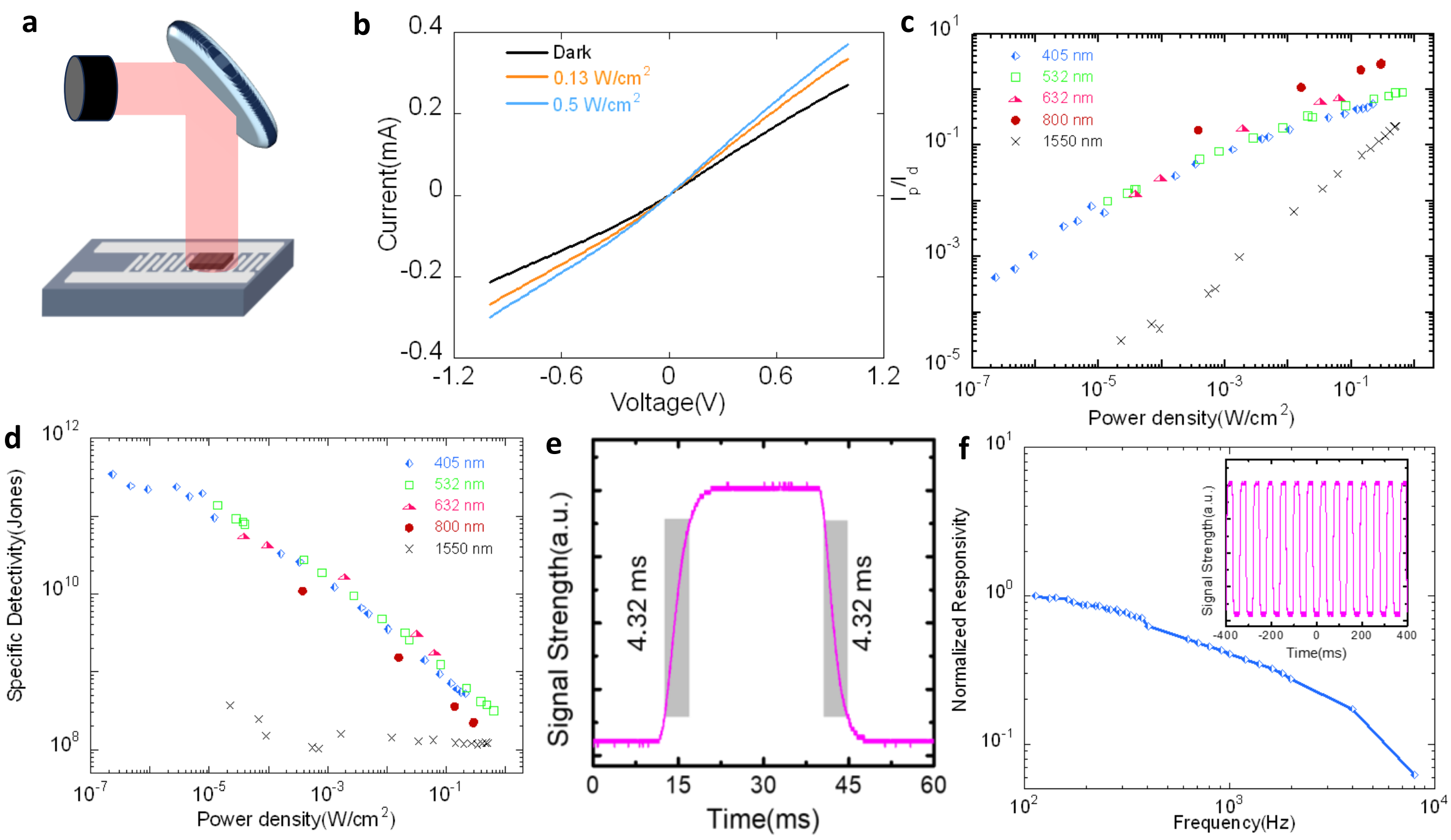

**Figure 4.** Horizontal faster device characteristics: **(a)** a schematic of a device described with light illumination from top after reflection from a silver mirror **(b)** Current-Voltage sweeps of the device are shown for three powers of illumination of 1550 nm source including the no illumination curve. **(c)** Ratio of photocurrent and dark current $(I_p/I_d)$ is shown to be linearly varying with power density of laser sources i.e. 405 nm, 532 nm, 632 nm, 808 nm and 1550 nm. **(d)** Specific detectivity varying with illumination power of mentioned wavelengths **(e)** Rise time and fall time of the device is calculated to 4.32 ms for a device under 1550 nm illumination. **(f)** Normalized responsivity of the device when chopped the 1550 nm source of illumination with chopping frequency up to 8 kHz range. Repetitive light on/off cycles are shown in inset.

The aged devices exhibit stable and reproducible photoconductivity throughout the investigated optical power range. Representative current–voltage characteristics recorded under 1550 nm illumination are presented in Figure 4b, showing a monotonic increase in photocurrent with increasing optical intensity. Additional measurements at different optical powers are shown in Figure S12.

The dependence of photocurrent on optical power reveals two distinct transport regimes. Under visible illumination, the normalized photocurrent follows approximately ($I_p/I_d \sim P^{1/2}$), whereas excitation at 1550 nm produces an almost linear dependence, ($I_p/I_d \sim P^1$) (Figure 4c and Figure S13). Sublinear behaviour under visible illumination is characteristic of selenium-based photoconductors[30] and is generally associated with trap-assisted recombination.[31] In contrast, the nearly linear response observed at 1550 nm indicates efficient carrier generation with comparatively weaker trap-limited transport, suggesting that structural relaxation substantially suppresses carrier trapping in the infrared regime.

The aged Se-$I_2$ photodetectors exhibit responsivities and specific detectivities comparable to previously reported selenium-based infrared photodetectors. Shot-noise-limited specific detectivities [32–34] reach approximately $10^8$ Jones at 1550 nm and $10^{11}$ Jones in the visible spectral region (Figure 4d), while the responsivity decreases gradually with increasing optical power (Figure S14a), consistent with saturation effects commonly observed in selenium photoconductors.[35–37] Although the photocurrent remains significantly smaller than the dark current over much of the investigated power range (Figure S14b), the photocurrent increases continuously with optical intensity and approaches the dark-current magnitude at the highest incident powers investigated.

The most striking consequence of structural relaxation is the dramatic improvement in temporal response. Following aging, the photodetector exhibits rise and decay times of approximately 4.3 ms under 1550 nm excitation (Figure 4e). Prior to aging, only a slow persistent photoconductive component was observed (Section 3.4). After structural relaxation, however, a fast millisecond-scale response emerges while the slow bolometric contribution remains as a secondary component. The fast response becomes particularly evident under higher modulation frequencies, where the persistent component can no longer follow the optical excitation (Figure 4e).

Comparison with the unaged devices reveals an improvement in response speed by nearly three orders of magnitude (Figure S15), demonstrating that structural disorder within the initially glassy Se-$I_2$ phase is the dominant factor limiting carrier transport. Similar improvements in electronic transport accompanying structural ordering have previously been reported in polymeric chain-like semiconductor systems, where increased structural coherence reduces carrier localization and facilitates more efficient charge transport.[38,39]

The improved transport characteristics also enable stable high-frequency operation. As shown in Figure 5f, the normalized photoresponse remains nearly constant up to modulation frequencies of approximately 8 kHz, with highly reproducible on/off switching (Figure 4f, inset), confirming robust device operation under repeated optical modulation. Furthermore, lowering the operating temperature substantially suppresses the dark current while simultaneously increasing the photocurrent-to-dark-current ratio (Figure S16), consistent with thermally activated dark conduction in selenium-based semiconductors[40,41]. These results collectively demonstrate that structural relaxation transforms Se-$I_2$ from a sluggish glassy photoconductor into a practical millisecond infrared photodetecto

## 3.6 Thermal Imaging and Line Scanning of the Device

### 3.6.1. Thermal Imaging:

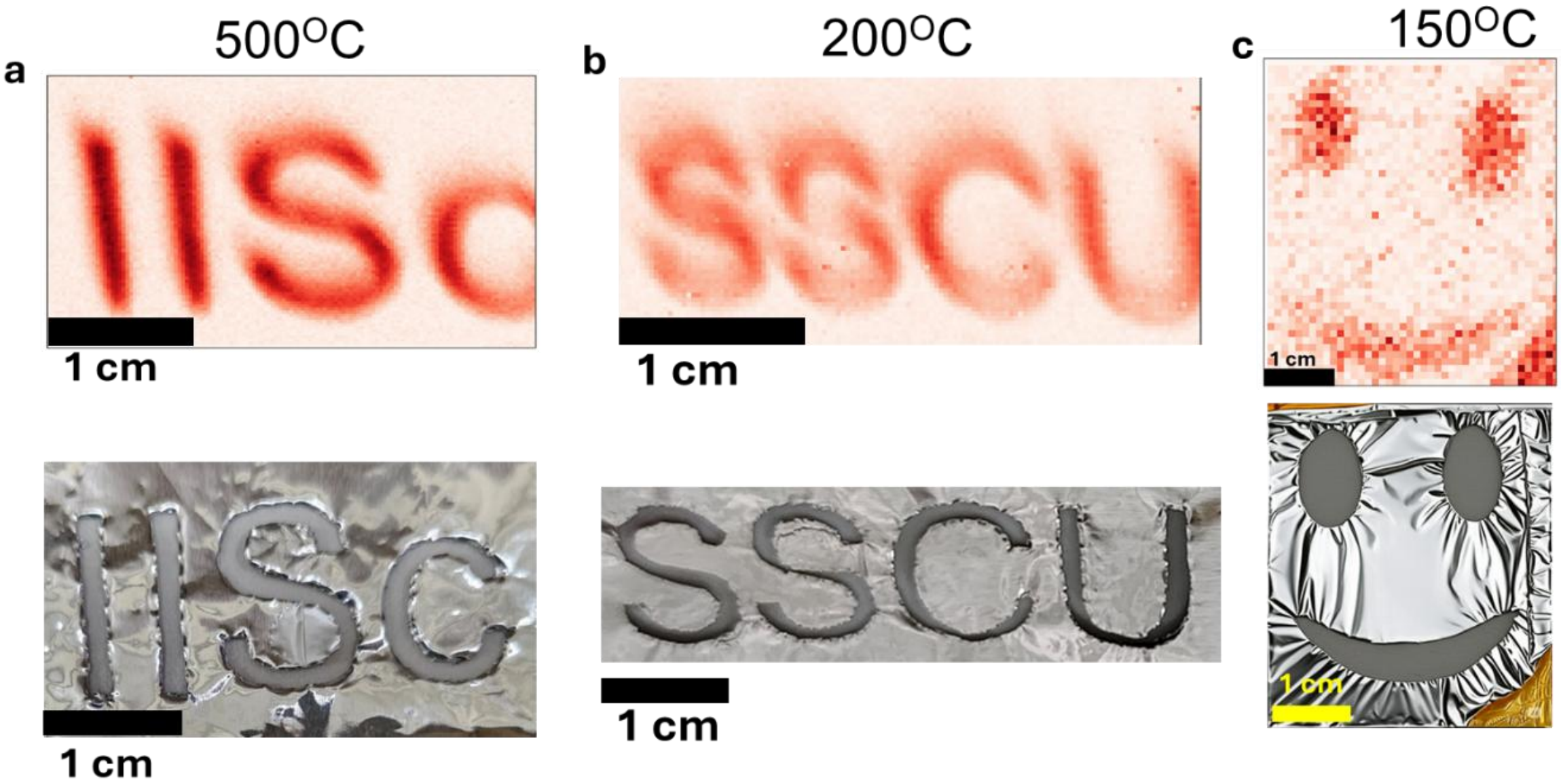

**Figure 5.** Imaging of  the device: **(a)** “IISc” and **(b)** “SSCU” and **(c)** the Smiley Shape was imaged at 500°C, 200°C and 150°C. The shapes are shown below the corresponding images are made by perforating the an aluminium foil in a desired shape. All the images are associated with the corresponding scale bar.

Having established the high-speed infrared photoresponse of structurally relaxed Se-$I_2$, we next evaluated its capability for practical thermal imaging. Infrared emission from heated objects was projected onto the photodetector surface using patterned aluminum masks placed over a ceramic hot plate (Supplementary Section 10). The exposed regions of the hot plate acted as localized infrared emitters, while the aluminum mask suppressed thermal emission from the surrounding areas, thereby generating high-contrast infrared patterns.

To demonstrate operation over a broad temperature range, three representative thermal patterns were investigated. The letters "IISc" were imaged using a source temperature of 500 °C (Figure 5a), the pattern "SSCU" was recorded at 200 °C (Figure 5b), and a smiley-face pattern was successfully reconstructed at only 150 °C (Figure 5c). These experiments demonstrate reliable infrared detection over temperatures spanning several hundred degrees, encompassing many practical thermal imaging scenarios.

Images were acquired by raster scanning the detector across the projected infrared field while recording the photocurrent at each position. The measured photocurrent was subsequently converted into spatial intensity maps representing the thermal emission profile. The reconstructed images faithfully reproduce the original object geometries with excellent spatial contrast, confirming both the sensitivity and spatial resolution of the Se-$I_2$ photodetector.

The successful detection of relatively weak thermal emission from the 150 °C source highlights the sensitivity of the detector in the near-infrared spectral region. This performance originates from the extended infrared absorption edge of Se-$I_2$ (Figure S18), which enables efficient absorption of long-wavelength thermal radiation inaccessible to elemental selenium.

These imaging experiments demonstrate that the improved transport characteristics achieved through structural relaxation translate directly into practical device functionality. Beyond conventional photoconductivity measurements, the ability to reconstruct spatially resolved infrared images establishes Se-$I_2$ as a viable semiconductor platform for infrared sensing. Considering its solution processability, low fabrication cost, and millisecond response time, Se-$I_2$ represents a promising alternative material for applications including thermal monitoring, industrial inspection, environmental sensing, and infrared imaging technologies.

### 3.6.2. Spatial Uniformity of the Photoresponse:

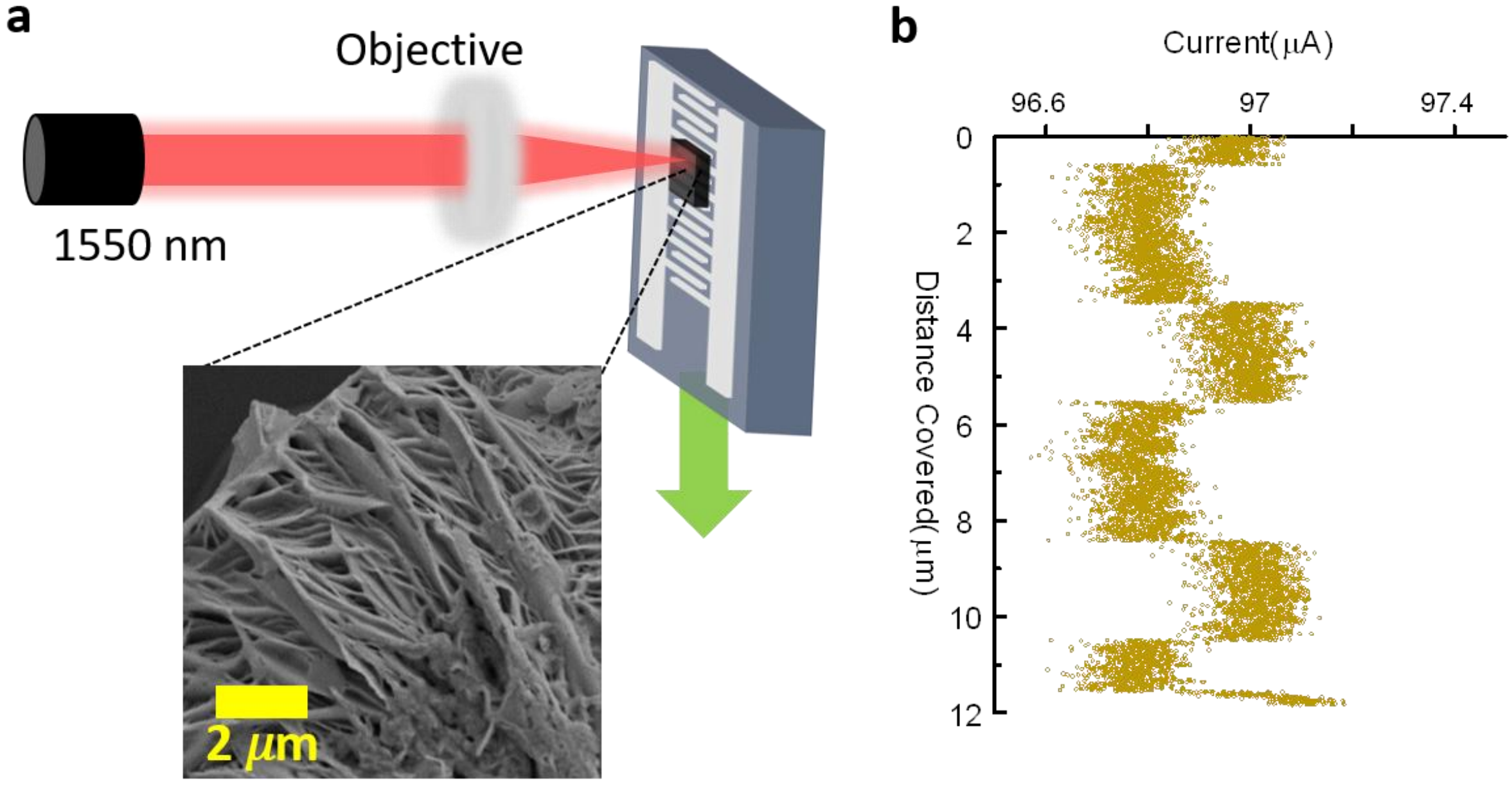

**Figure 6.** Line scanning of the horizontal device: **(a)** "Line scan of the horizontal device: a 1550 nm source is used in combination with an objective lens to make a tight spot on the film. For the line scanning of photo response, the device was moved vertically perpendicular to the beam path. The film SEM image is shown in the zoomed image. **(b)** The current is shown to rise and fall in every ~2 *μ*m distance it travels.

Spatially resolved photocurrent mapping was performed to evaluate the uniformity of carrier collection across the active area of the Se-$I_2$ photodetector, an important consideration for large-area infrared imaging applications.[42,43] The photocurrent was recorded under focused 1550 nm laser illumination while raster scanning the excitation spot across the interdigitated Pt electrode device (Figure 6a). The corresponding scanning electron microscopy image reveals that the Se-$I_2$ film consists of interconnected one-dimensional lamellar structures extending throughout the active region.

Despite the pronounced surface morphology, the spatial photocurrent map (Figure 6b) exhibits excellent uniformity, with only minor signal fluctuations of approximately 0.4% across the scanned area. These small variations are attributed primarily to local topographical variations associated with the lamellar microstructure rather than intrinsic electronic inhomogeneity. The absence of significant spatial variations in the photocurrent indicates efficient and homogeneous carrier generation and collection throughout the device[44,45], demonstrating that the self-assembled lamellar architecture does not compromise large-area optoelectronic performance. This high degree of spatial uniformity, together with the demonstrated millisecond response and infrared imaging capability, highlights the potential of Se-$I_2$ as a scalable semiconductor platform for practical infrared photodetection.

# 4. Conclusion:

In summary, we demonstrate that the optoelectronic performance of selenium-iodine (Se-$I_2$) is intrinsically governed by its structural evolution during solidification and subsequent aging. While freshly solidified Se-$I_2$ forms a metastable glassy structure that exhibits sluggish carrier transport and persistent photoconductivity, gradual structural relaxation into a more ordered lamellar architecture dramatically accelerates carrier dynamics, enabling millisecond-scale infrared photodetection. Through vertically resolved transport measurements, we further identify pronounced depth-dependent electrical conductivity within thick Se-$I_2$ films, providing direct evidence that compositional and structural inhomogeneity critically influence charge transport in this material.

The structurally relaxed devices exhibit stable infrared photodetection at 1550 nm with millisecond response times, reproducible operation at modulation frequencies up to 8 kHz, and shot-noise-limited detectivities comparable to those of reported selenium-based infrared photodetectors. Beyond conventional device characterization, we demonstrate spatially resolved infrared thermal imaging together with highly uniform photocurrent generation across the active device area, highlighting the robustness of the self-assembled lamellar Se–$I_2$ architecture for practical imaging applications.

More broadly, this work establishes the intimate relationship between structural ordering and optoelectronic functionality in selenium-iodine semiconductors and demonstrates that controlling structural relaxation provides an effective route for engineering carrier transport and photodetection performance. These findings position iodine-modified selenium as a promising solution-processable semiconductor platform for next-generation infrared photodetectors and thermal imaging technologies while providing broader insight into the design of structurally evolving semiconductor systems.

**5. Author Contribution:** B.J. set up experiments, collected data and wrote the manuscript.

**6. Acknowledgements:** B.J. acknowledge funding support from Govt. of India through Prime Research Fellowship and Research Grant with project code: PM/MHRD-19-14871.03. B.J. acknowledge Anshu Pandey and Sushobhan Avasthi for insightful discussion. B.J. acknowledge Hridesh Kumar Gupta, for technical assistance I-V measurements and Ankur Bhaumik for thermal imaging. The authors acknowledge the use of ChatGPT (OpenAI) for language editing and improving the clarity of the manuscript. The authors critically reviewed, edited, and verified all generated content and assume full responsibility for the final manuscript.

# Structural Relaxation Enables Millisecond Infrared Photodetection in Selenium–Iodine Semiconductors

Biman Jana[1]*

[1]Indian Institute of Science, Bangalore, 560012, India

*Corresponding author, *Email:* biman.jana013@gmail.com, bimanjana@iisc.ac.in

## 1. Dark Current and Photocurrent Variation upon Crystallization:

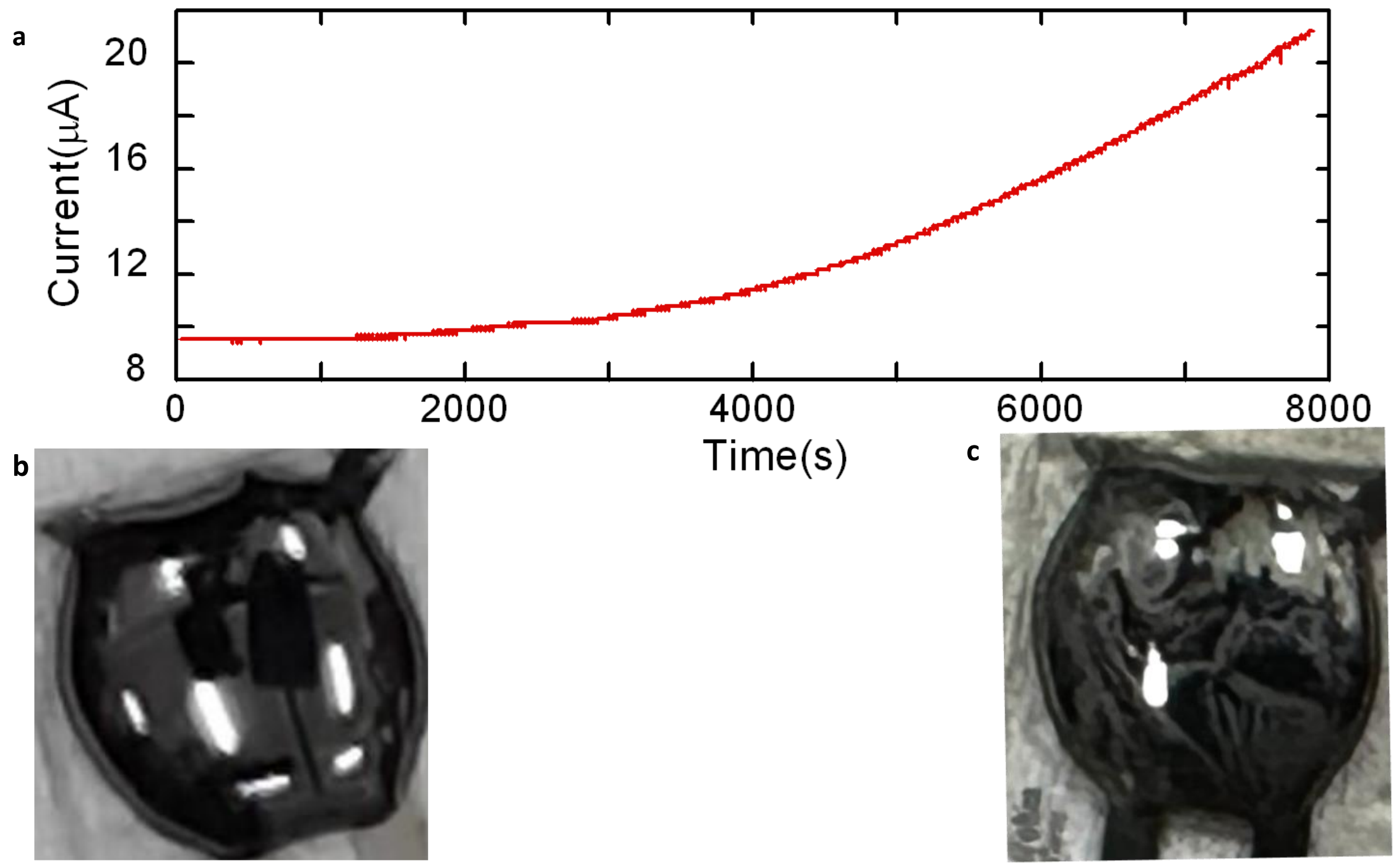

**Figure S1:** Dark current variation with crystallization of the material. The picture on the left shows the Se-$I_2$ material in liquid state. The picture on the right shows the material after allowing it to crystallize for 2 hour.

The dark current flowing through the material is probed with two platinum electrode deposited on glass slide. As evident from Figure S1, the dark current improves with time when the material start to recover its structural order.

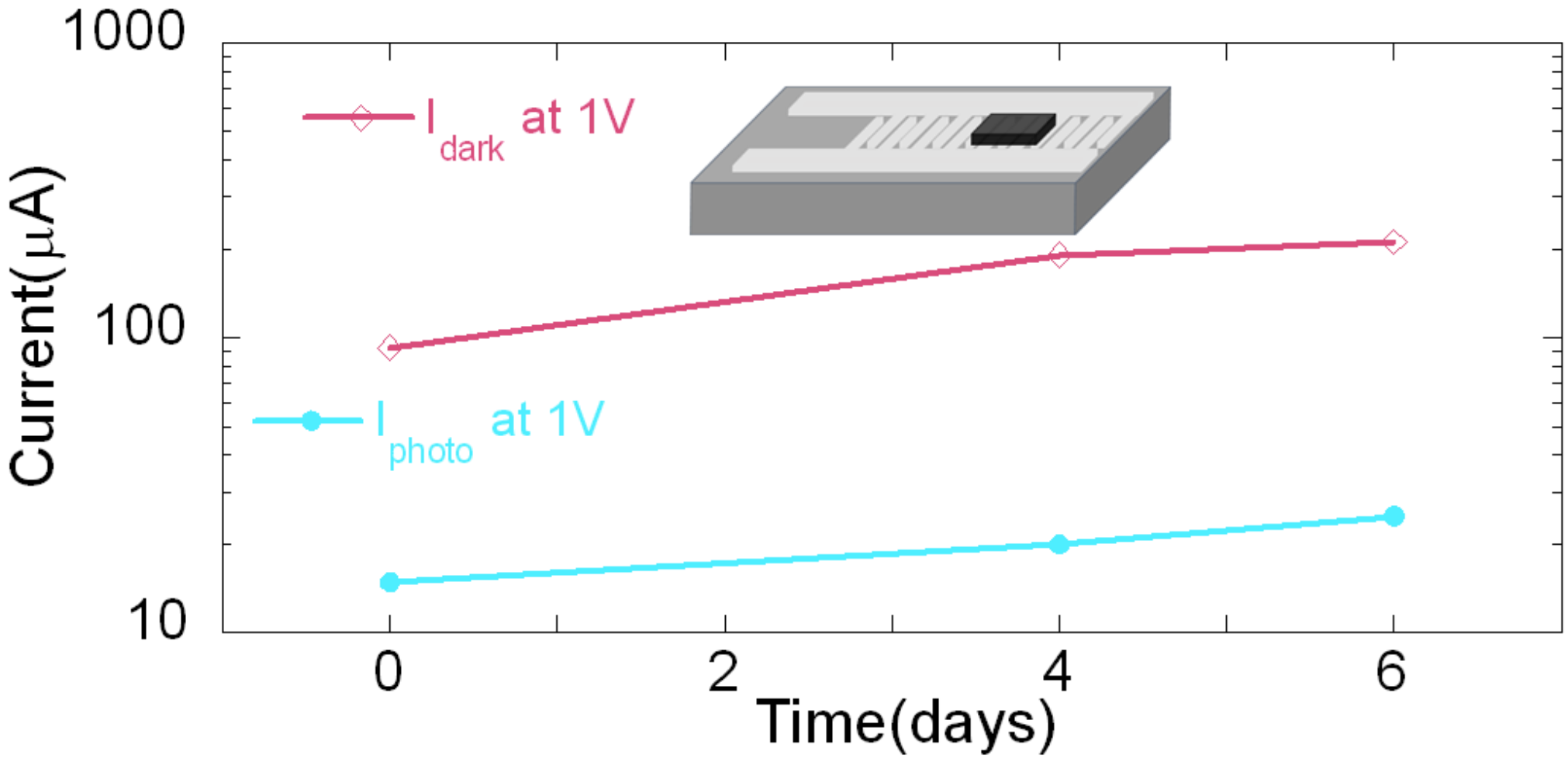

**Figure S2:** Variation of dark current and photocurrent with time for a typical device made by depositing $Se_nI_2$ on platinum interdigitated electrode with 10 $\mu$m width and 10 $\mu$m separation between the electrodes as depicted in the inset. The photoresponse in $0^{th}$ days was taken after allowing the liquid to solidify for 12 hours. The dark current and photocurrent is seen to be increasing with time.

## 2. Structural Characterizations:

### 2.1 SEM Image and Elemental Mapping:

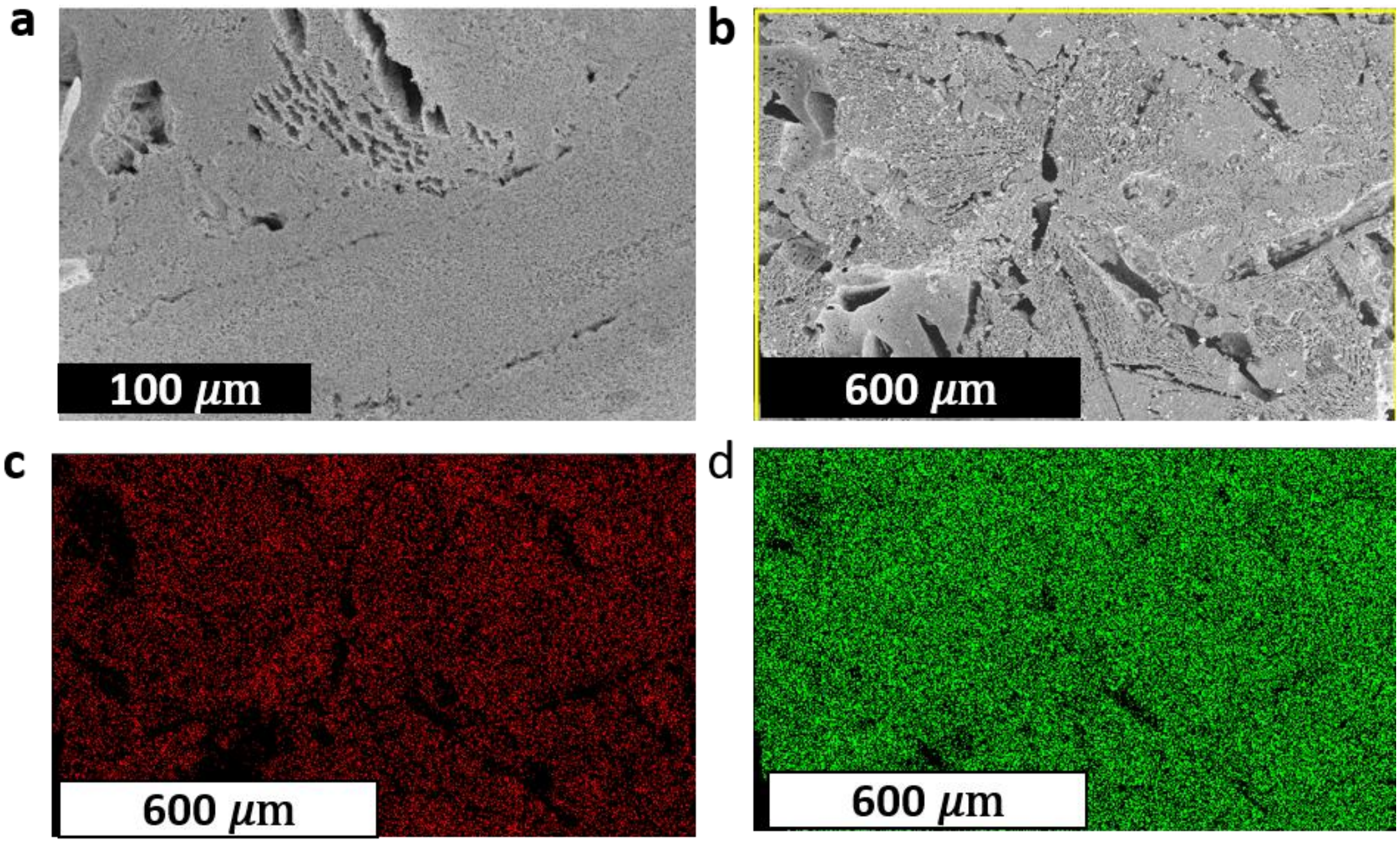

**Figure S3: (a-b)** SEM image of typical film on silicon and **(c) and (d)** energy dispersive X-ray spectra EDS elemental mapping for selenium and iodine respectively for the area shown in **(b)**.

The porous structure of the Se-$I_2$ is obvious from the SEM image (Figure S4a-b). The individual characteristic X-ray emission is identified for both the elements through Energy dispersive X-ray spectroscopy (EDS) elemental mapping. Uniform distribution of selenium and iodine across the horizontal dimension is obvious from Figure S4c-d except the voids.

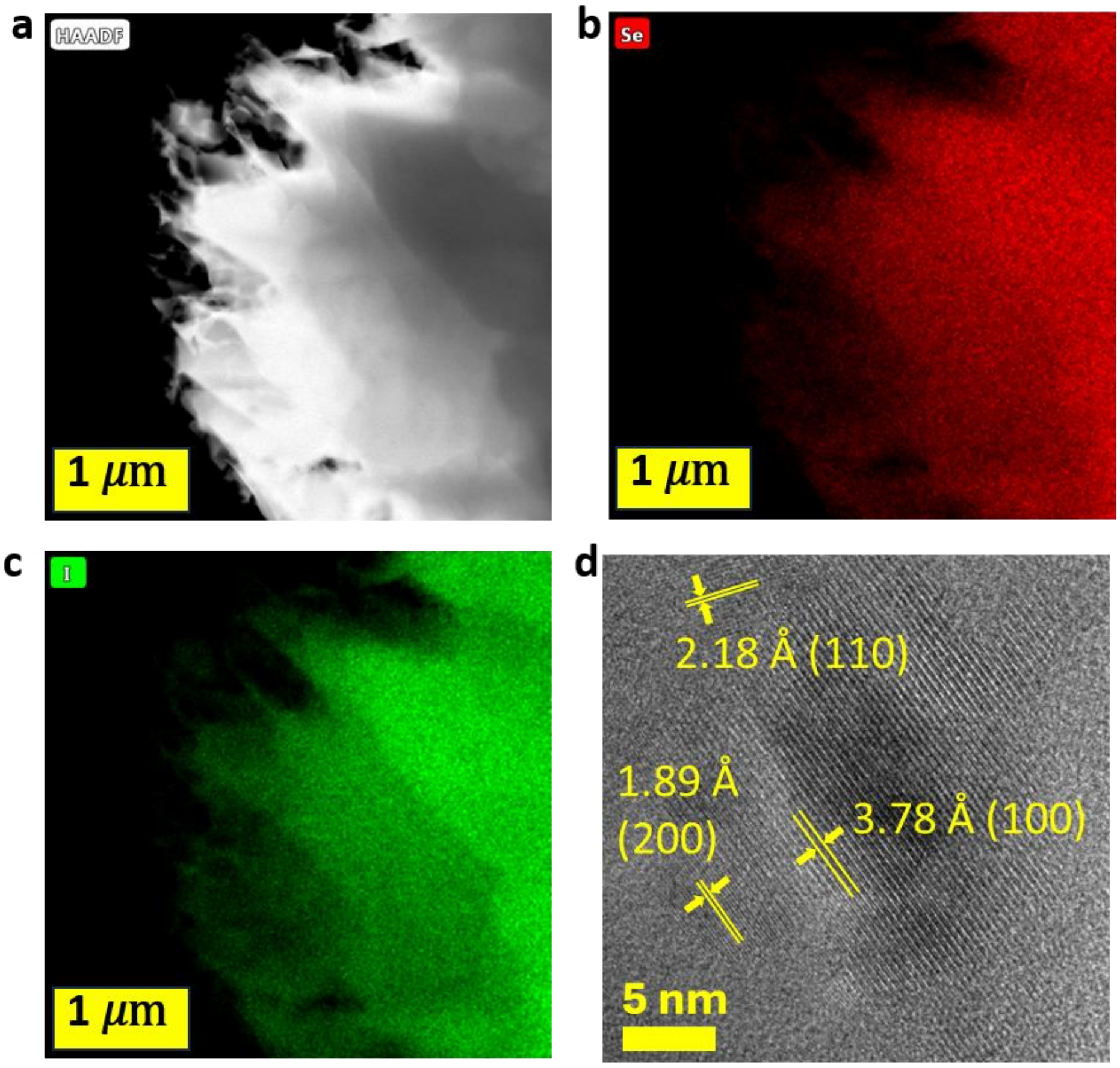

**Figure S4**: **(a)** High angle annular dark field-scanning transmission electron microscopy image and corresponding elemntal mapping for the **(b)** Se and **(c)** iodine of the Se-$I_2$. HRTEM analysis marking trigonal selenium planes.

Because molecular iodine ($I_2$) is a subliming molecular solid with high vapor pressure, lattice planes from iodine crystals are difficult to observe during HRTEM experiments. Electron-beam irradiation can induce localized heating and radiolytic effects, causing the iodine crystals to sublime or decompose which prevents stable high-resolution imaging.

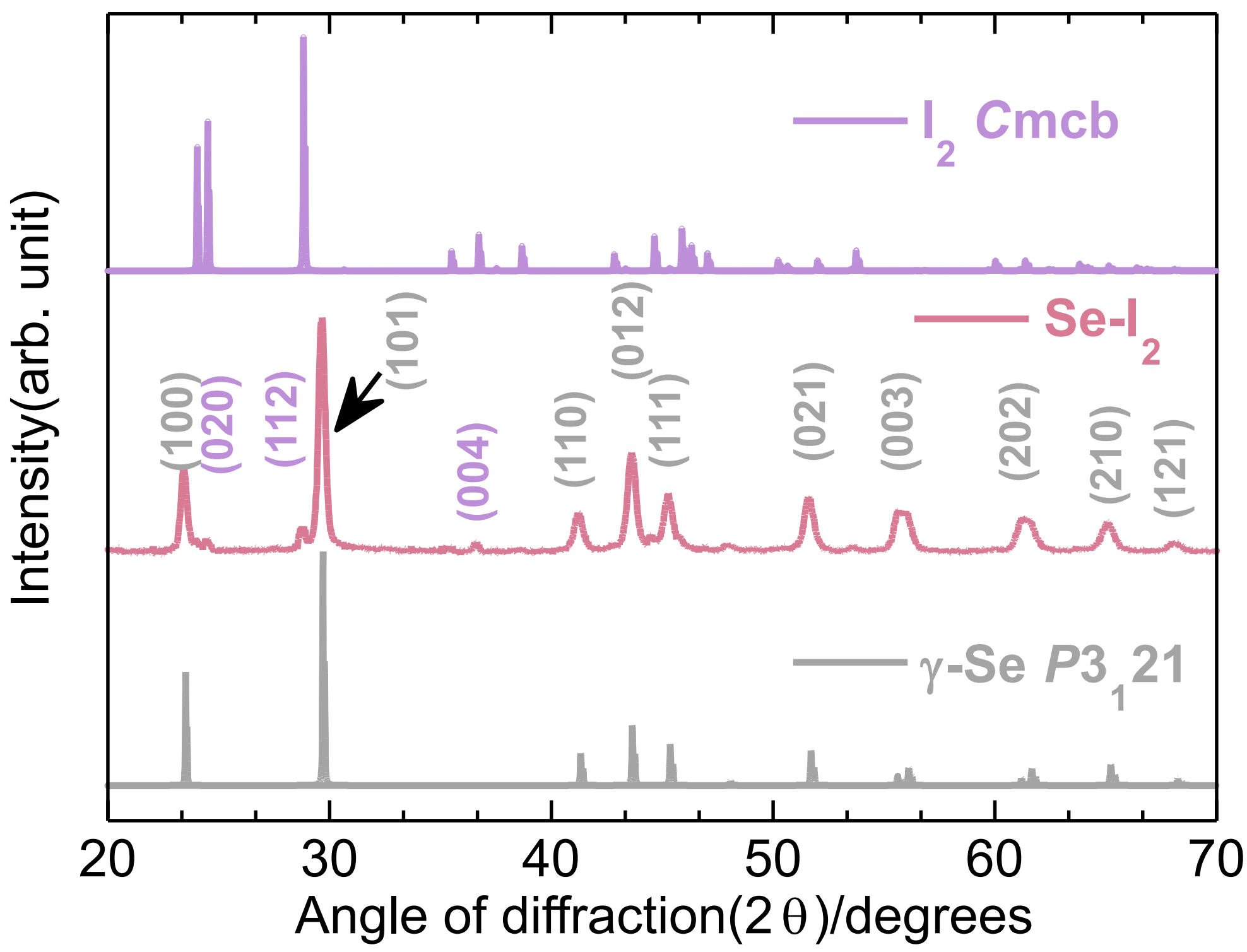

**Figure S5**: X-ray diffraction pattern of solidified Se-$I_2$ om glass substrate collected with Cu $K_\alpha$ source with a PAnalytical-Empyrean instrument.

## 3. Preparing Film of Se-$I_2$ For Vertical Devices:

64 mg of Selenium powder and 200 mg of molecular Iodine was mixed and heated up in a closed argon environment. The mixture was then immediately decanted on the substrate. A Si wafer was placed on top of the liquid to ensure a flat smooth surface. The film was allowed to solidify for one day before making the successive layer on top. Platinum contacts are made by sputtering a 150 nm platinum layer using a mask on top with an ANELVA RF sputtering machine.

## 3.1. Difficulty in Probing with Only Metal Layer on Top of Porous Surface of Se-$I_2$:

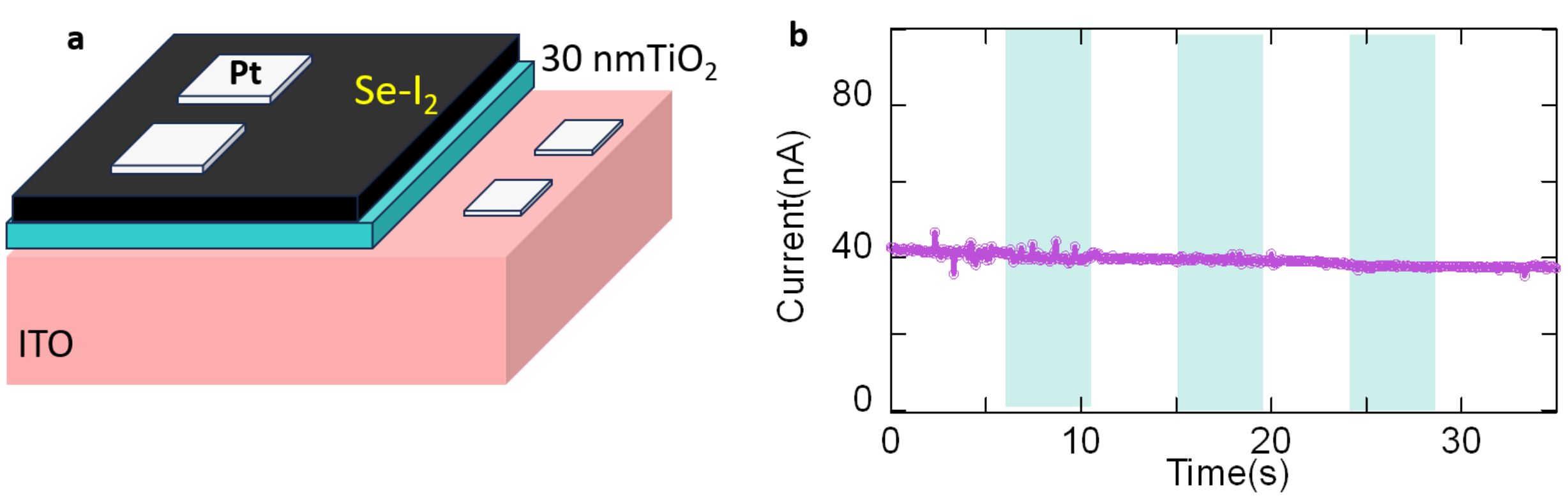

**Figure S6:** **(a)** Device geometry and **(b)** absence of photo response and unstable current due to difficulty in probing. The green line represents illumination with 1550 nm source.

Due to porous surface of the $SenI_2$, the sputtered metal forms a non-uniform layer on top of $SenI_2$ surface. This causes difficulty in probing which leads to unstable current and absence of photoresponse.

## 5. Vertical Device Response:

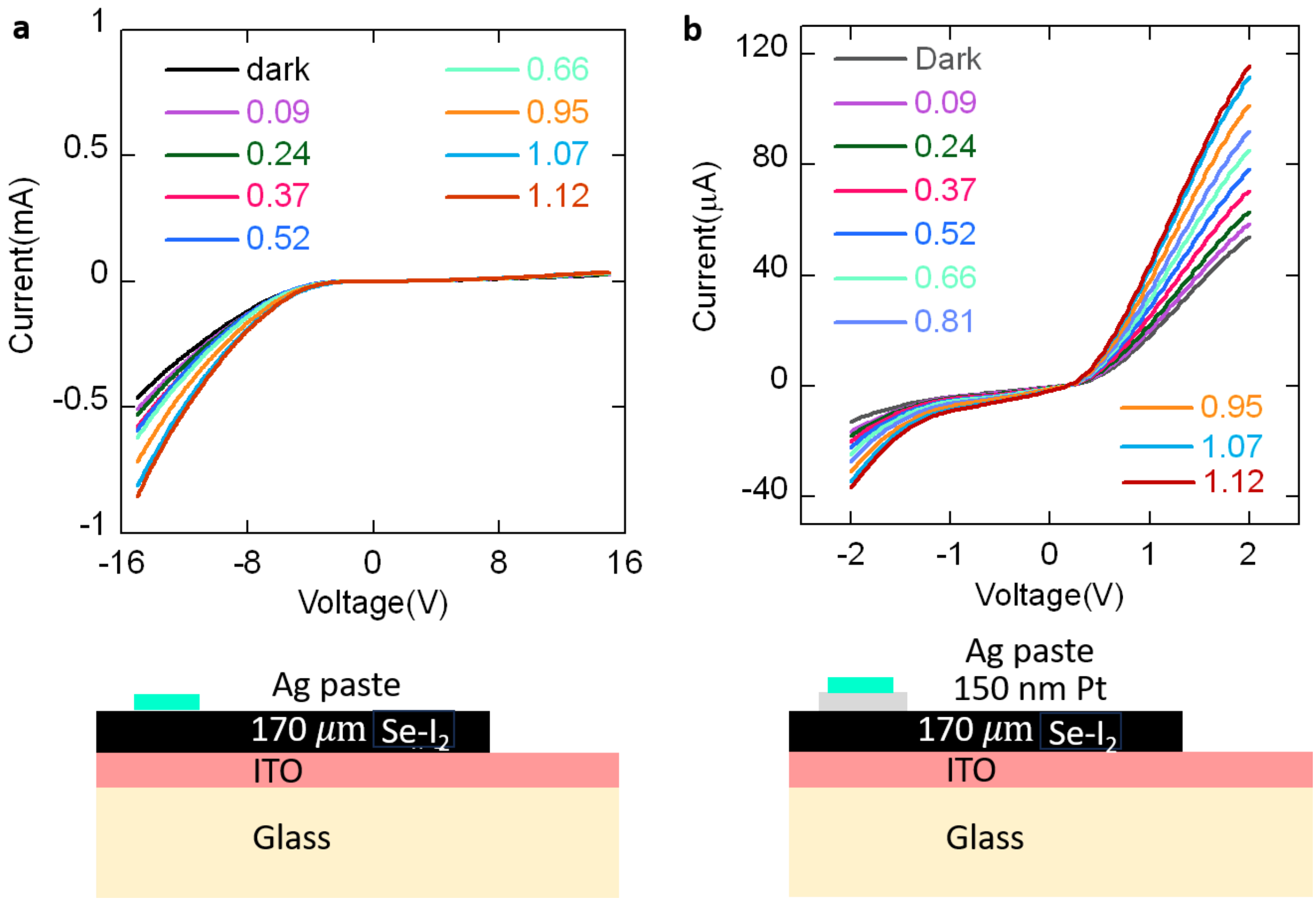

**Figure S7:** Comparison of two vertical devices: I-V characteristics of vertical stack device where **(a)** Ag paste is used directly on top of Se-$I_2$ layer and **(b)** 150 nm Platinum layer is used before making electrical contact with Ag paste on top of it.

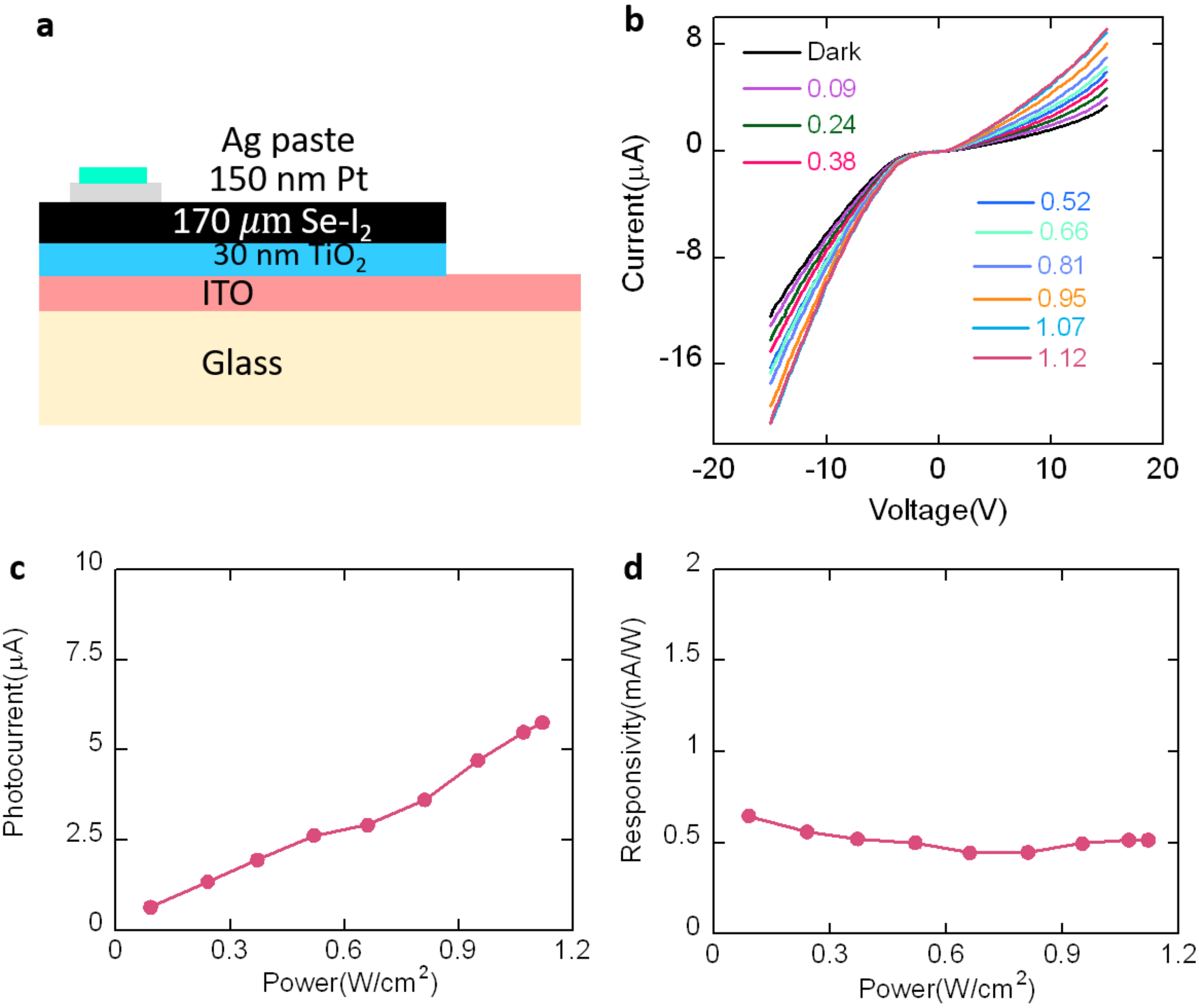

**Figure S8: (a)** Vertical stack device geometry designed with an extra layer of Ti2 as an electron transport layer. **(b)** I-V characteristics of vertical stack device (ITO/$TiO_2$/$Se_nI_2$/Pt/Ag). Here an extra $TiO_2$ layer is used as electron transport layer in contact with the active layer of $Se_nI_2$. **(c-d)** Photocurrent and responsivity is plotted against power of 155o m illumination at 15 V of bias applied

# 5. Photodetector device parameters:

**5.1. Photocurrent to Dark Current Ratio:** The rise in current flowing through a photodetector upon excitation with light is called the photocurrent

$$\frac{I_{photo}}{I_{dark}} = \frac{I_{bright} - I_{dark}}{I_{dark}} \qquad (1)$$

Where $I_{photo}$ is the photocurrent and $I_{bright}$ is the current seen in the circuit under an applied bias after exposing it to a light of particular intensity and wavelength. $I_{dark}$ is the dark current flowing through the device (without any light illumination). The ratio of photocurrent to dark current is a measure of the number of carriers generated by light excitation compared to the number of carriers already existing in the semiconductor.

**5.2. Responsivity:** Responsivity is the ratio of the output photocurrent to the incident optical power.

$R = \frac{I_{light} - I_{dark}}{P_{optical}}$ ……………………………………………………..……………………….(7)

This is a useful quantity to compare different detector performances. However, the noise current is not included here. This is why a more important term to compare different detectors is detectivity

**5.3. Detectivity:** Detectivity predicts how well a photodetector can detect weak signals in the presence of noise.

$D^* = \frac{Responsivity}{noise\ current}\sqrt{Area}$ ……………………………………………………….……….…….………(8)

With the assumption that the device is shot noise limited,

The e formula is reduced to $D^* = \frac{Responsivity}{\sqrt{2e(dark\ current)}}\sqrt{Area}$ …………………………………..(9)

For universal comparison of different photodetector of different sizes, materials, and time responses, detectivity is a useful quantity. the noise current in the denominator says that the lower the noise, the higher the detectivity.

# 6. Horizontal Slow Device Characteristics:

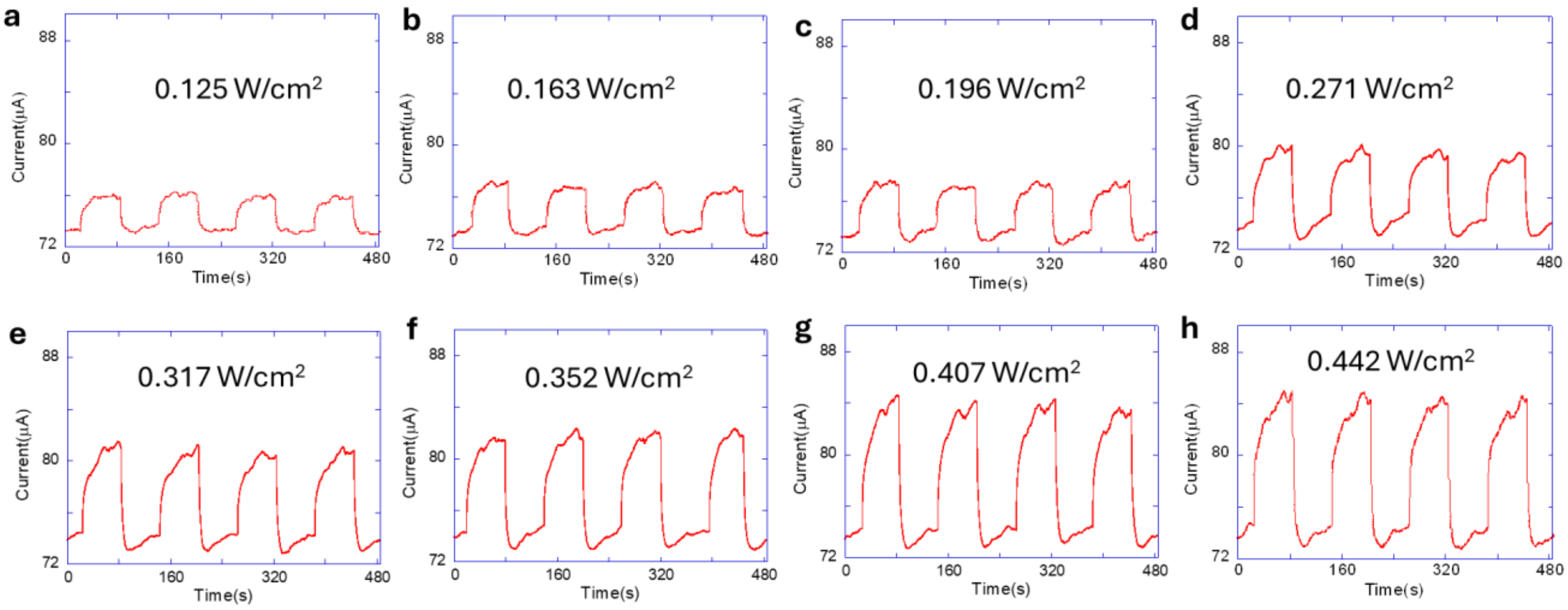

**Figure S9:** Photo response of the large area horizontal devices (5 mm × 7 mm) covering the at different power illumination of 1550 nm source as indicated in the Figures **(a-h).** The background has been subtracted to omit the continuous drift in current

## 6.1 Comparison of horizontal slow $Se_nI_2$ device with Mercury Cadmium Telluride detector:

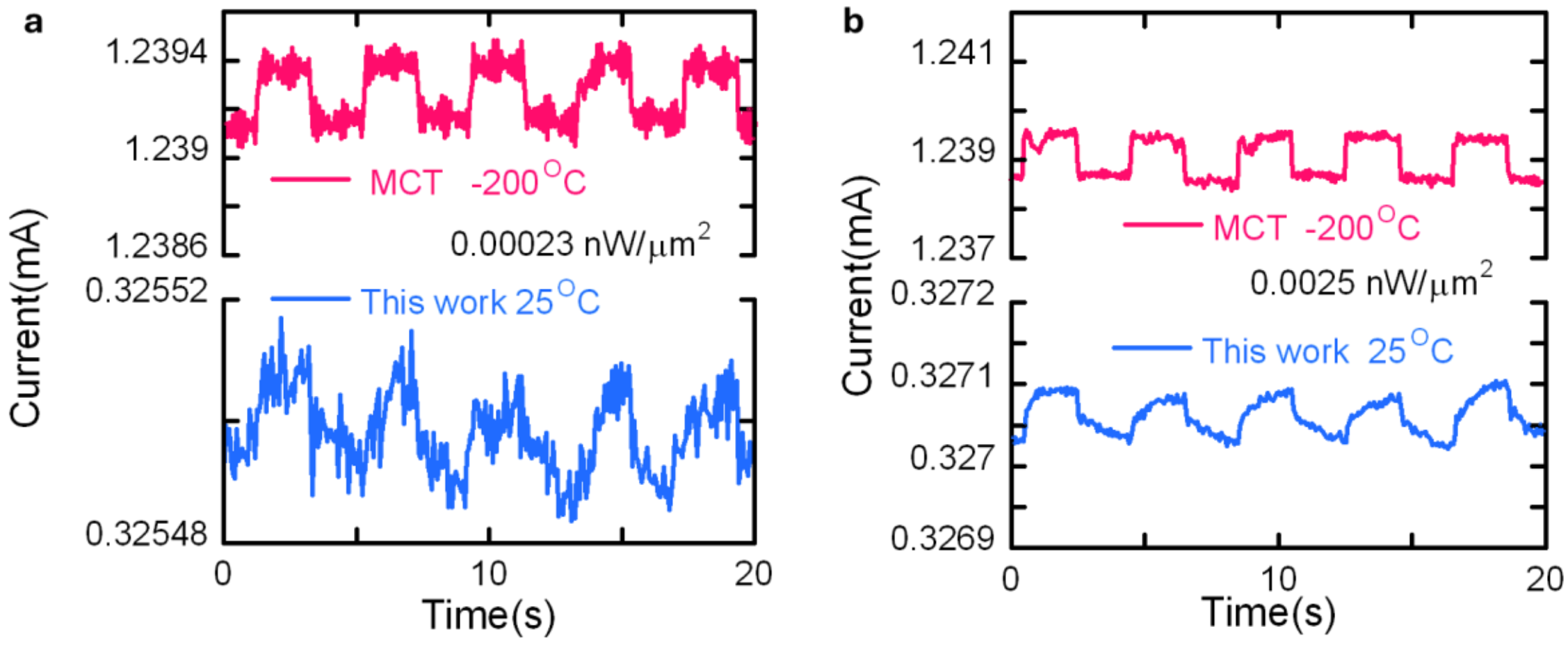

**Figure S10:** Comparison of device speed with respect to commercial Mercury cadmium telluride. the blue curve represents device response at 0.2 Hz of light chopping for 0.00023 nW/$\mu$m$^2$ and 0.0025 nW/$\mu$m$^2$. This shows the comparative result with available state of the art photodetector.

Mercury cadmium telluride ($Hg_{1-x}Cd_xTe$, MCT) detectors are currently considered the state-of-the-art in the field of infrared detection owing to their high quantum efficiency, fast response time (~1 ns), and tunable band gap in the 1–25 μm wavelength range depending on composition[1]. Commercially available MCT detectors typically exhibit nanosecond–microsecond response times and signal-to-noise ratios that are significantly higher than most emerging low-dimensional or chalcogenide-based materials. In contrast, the $Se_nI_2$ device studied here demonstrates much slower response speeds, particularly evident in the chopped light experiment at 0.2 Hz modulation frequency. While the blue curves in Figure 4.8 confirm that the device is photosensitive even at very low power densities (0.00023 nW/μm² and 0.0025 nW/μm²), the temporal response is markedly slower than the sharp response of the MCT detector.

It is important to note that the absolute vertical amplitude of the response in Figure 4.8a and 4.8b cannot be directly compared between the two systems because the operating voltages and electronic amplification conditions are not identical. Nevertheless, qualitative comparison of the rise and decay times, as well as the overall signal-to-noise characteristics, indicates that $Se\text{-}I_2$ is far inferior in both speed and noise performance. Specifically, the $Se\text{-}I_2$ device exhibits a broadened and delayed response tail. In contrast, MCT devices are known to exhibit nearly trap-free collection efficiency under optimized doping and growth conditions, which accounts for their superior temporal dynamics.

Finally, while MCT remains the benchmark for high-performance mid-IR detection, it is also limited by high fabrication costs, requirement of cryogenic cooling for best performance, and non-uniformity challenges over large-area growth. In this context, emerging low-dimensional chalcogenide systems such as $Se\text{-}I_2$ remain of significant research interest due to their potential for low-cost, flexible, and broadband detection, although at the expense of current performance limitations.

# 7. Characterization of the Faster Horizontal Device:

## 7.1 Thickness of the Device:

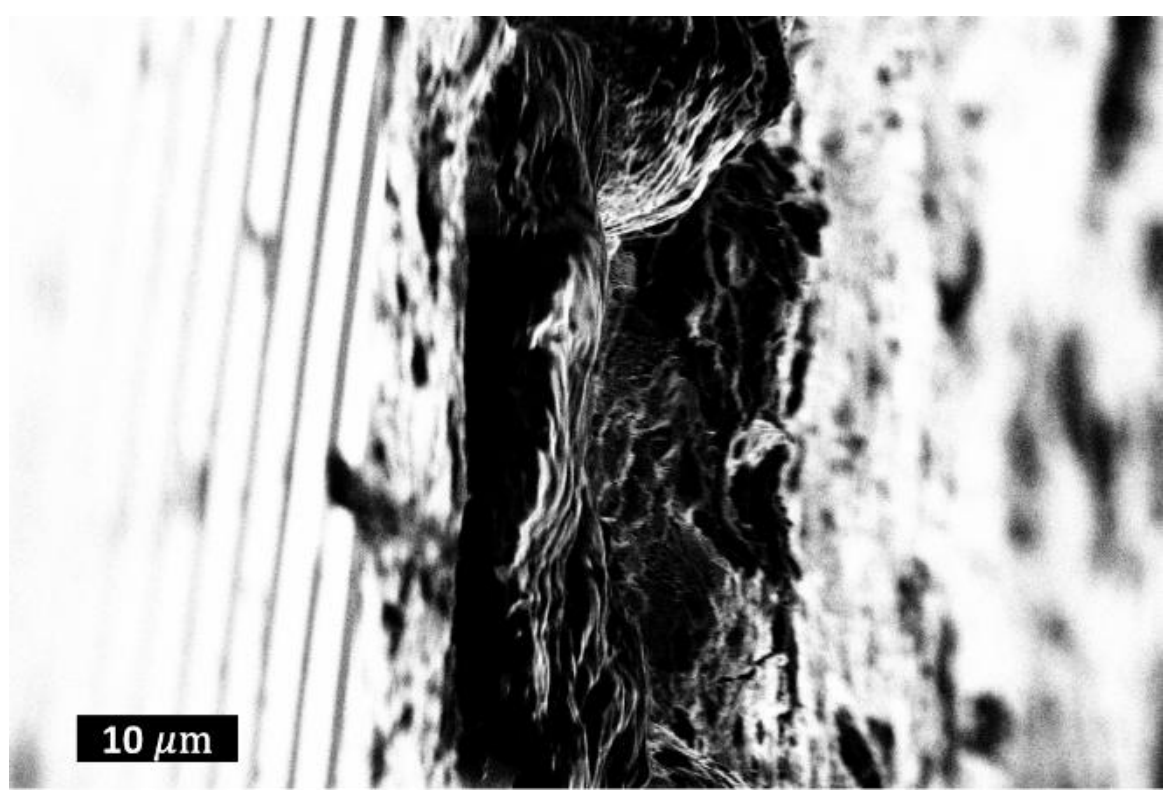

**Figure S11**: Tilted SEM image showing the thickness of the film ~17 $\mu$m

## 7.2 I-V Sweeps of Horizontal Device:

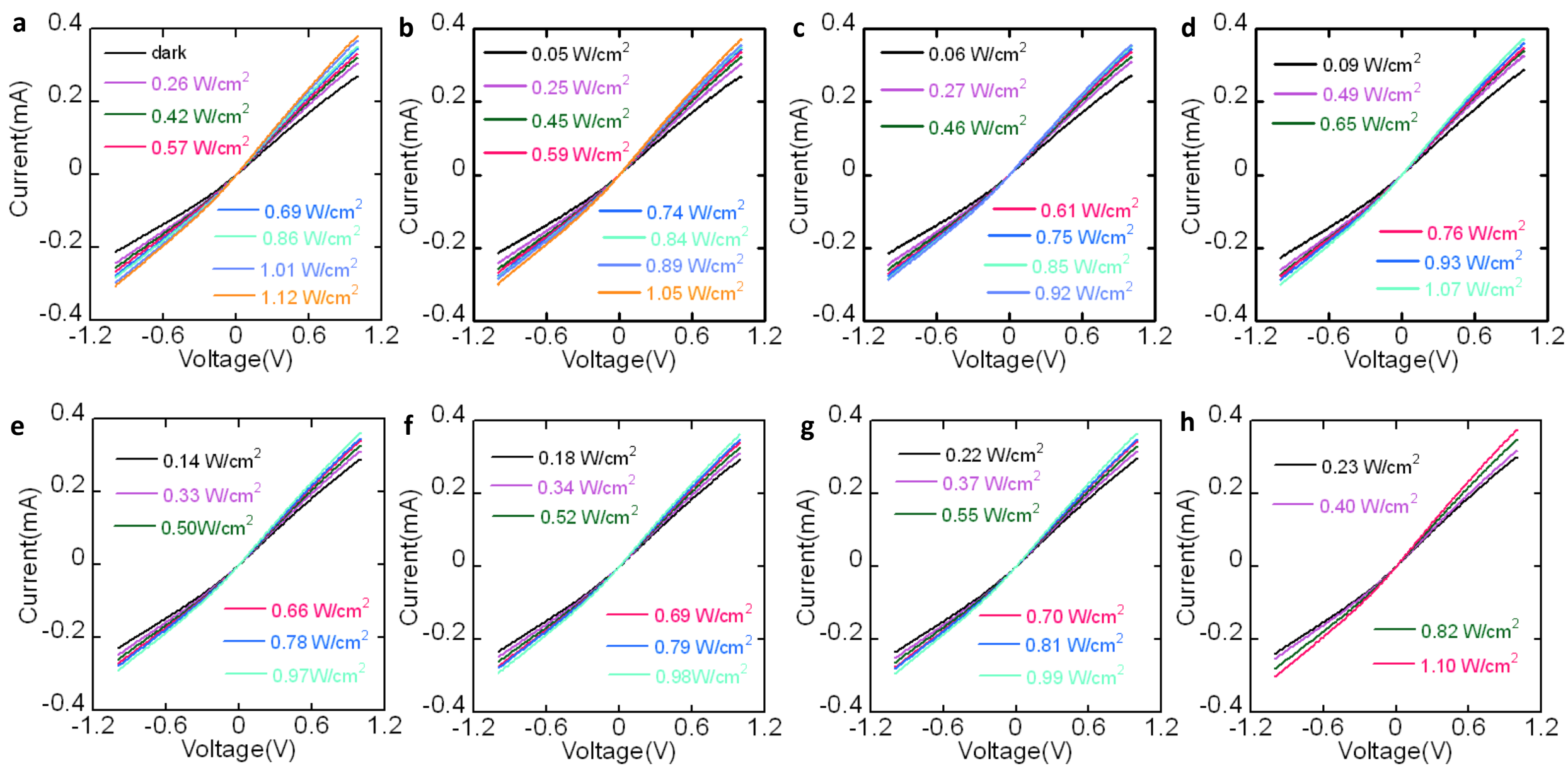

**Figure S12:** I-V characteristics of a typical horizontal Se-$I_2$ device with different power of illumination with 1550 nm source. The power densities are mentioned in legends.

## 7.3. Power Dependence of Photocurrent for Faster Device:

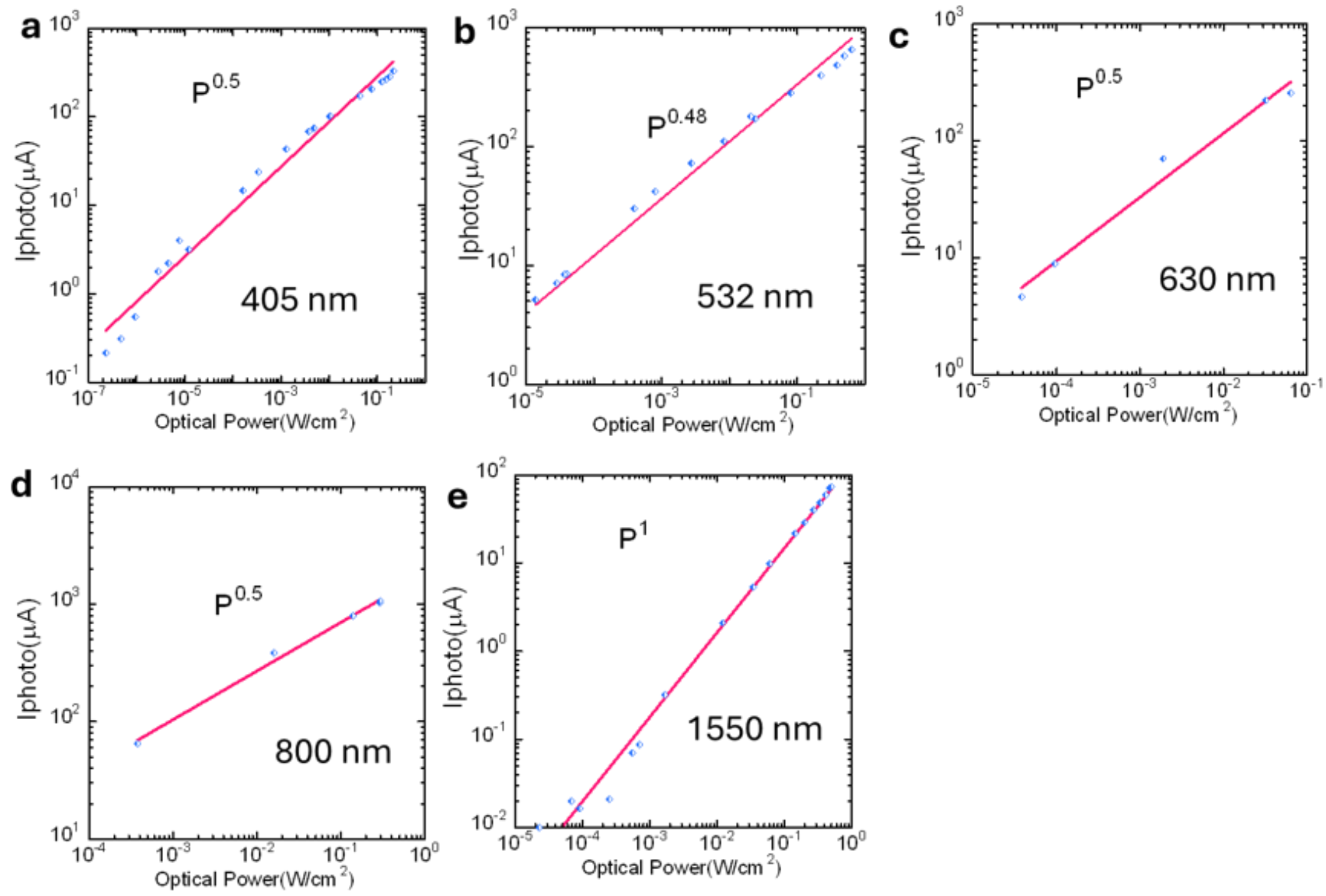

**Figure S13:** Optical power dependence of photocurrent **(a-e)** Photocurrent dependence of incident optical power density for 405 nm, 532 nm, 630 nm, 800 nm and 1550 nm. Up to 800 nm photocurrent varies as $\sim P^{0.5}$ and for 1550 nm the photocurrent varies as ~P. **(f)** Schematic to demonstrate the carrier recombination to explain the power dependence

After quantitatively fitting the photocurrent response as a function of incident optical power (Figure 4.11a–e), it was observed that the photocurrent increases monotonically with illumination power across all tested wavelengths. The dependence can be well described by the empirical power law relation:

$$I_p \propto P^n$$

where $I_p$ is the photocurrent, P is the incident optical power, and n is the power-law exponent that reflects the underlying recombination dynamics of photogenerated carriers.

For all visible wavelengths investigated, ranging from 405 nm to 808 nm, the exponent n is found to be close to 0.5. A sub-unity exponent of this magnitude suggests that the photocurrent is not simply proportional to the number of absorbed photons but is limited by non-linear recombination pathways. Specifically, an exponent around 0.5 is generally associated with bimolecular recombination processes. Interestingly, when probed at the telecommunication wavelength of 1550 nm, the photocurrent follows a nearly linear dependence on power, i.e. $I_p \propto P$ , where n = 1. This implies that in the lower-photon-energy regime, the carrier generation process becomes more efficient and largely free from bimolecular recombination bottlenecks. Here, each absorbed photon contributes effectively to the photocurrent without significant competition from recombination.

## 7.4. Horizontal Device Responsivity Varying with Optical Power:

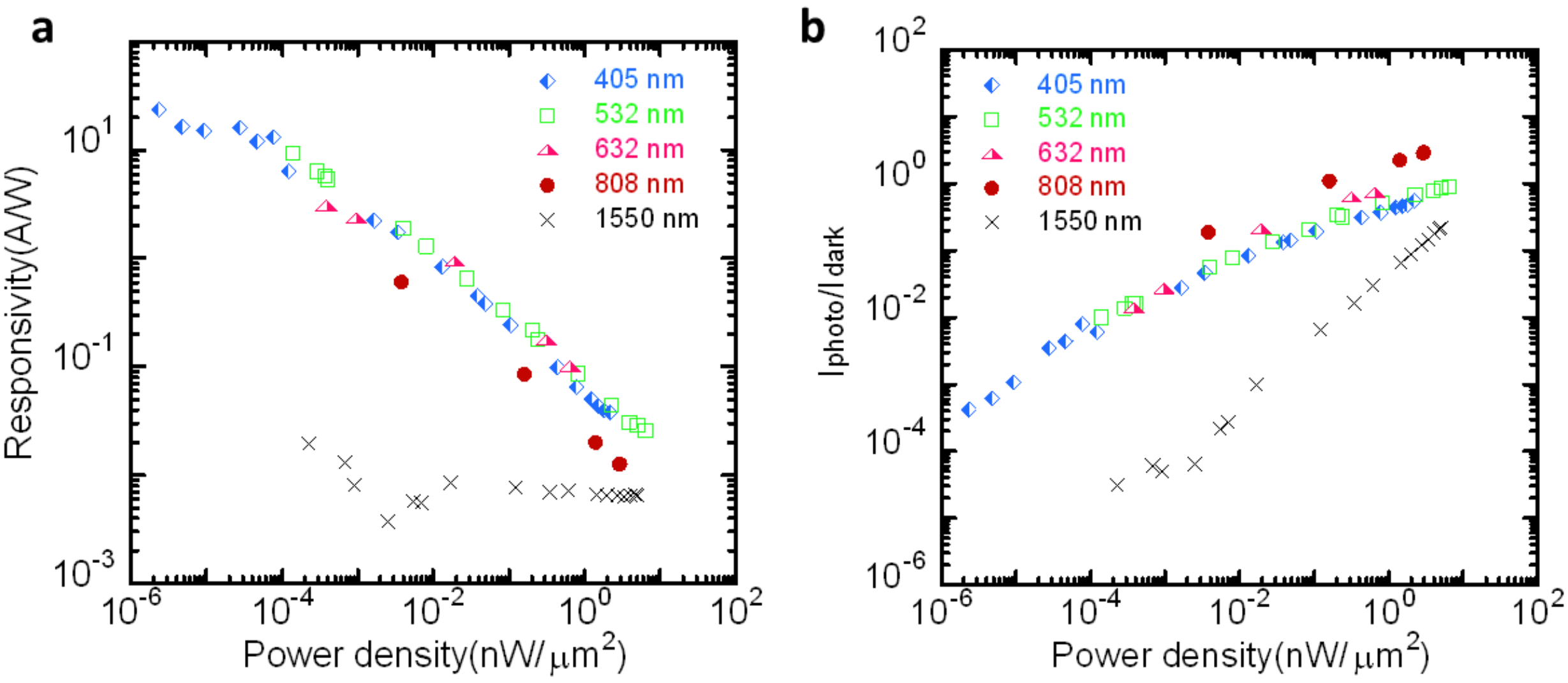

**Figure S14: (a)** Responsivity, **(b)** $I_{photo}/I_{dark}$ and variation when the device was illuminated with lights of different wavelengths for the horizontal fast photodetector presented in Figure 5

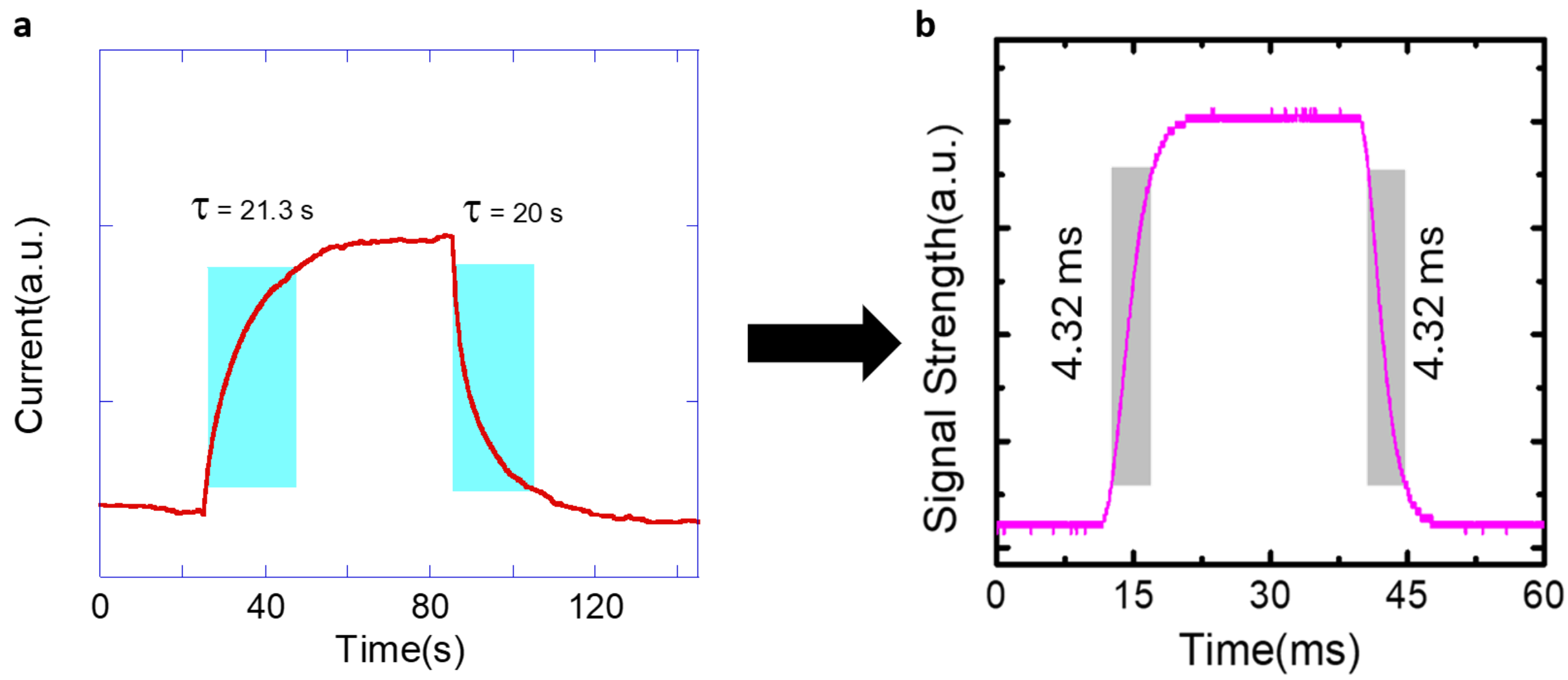

**Figure S15: (a)** Sluggish rise and fall on exposure and removal of a 1550 nm light source for slow vertical device. **(b)** Response time evaluation for optimized device showing ~5000 times faster response after removing the intercalated iodine and ageing in horizontal devices.

## 9. Low Temperature Photoresponse and I-V Sweeps at Different Powers:

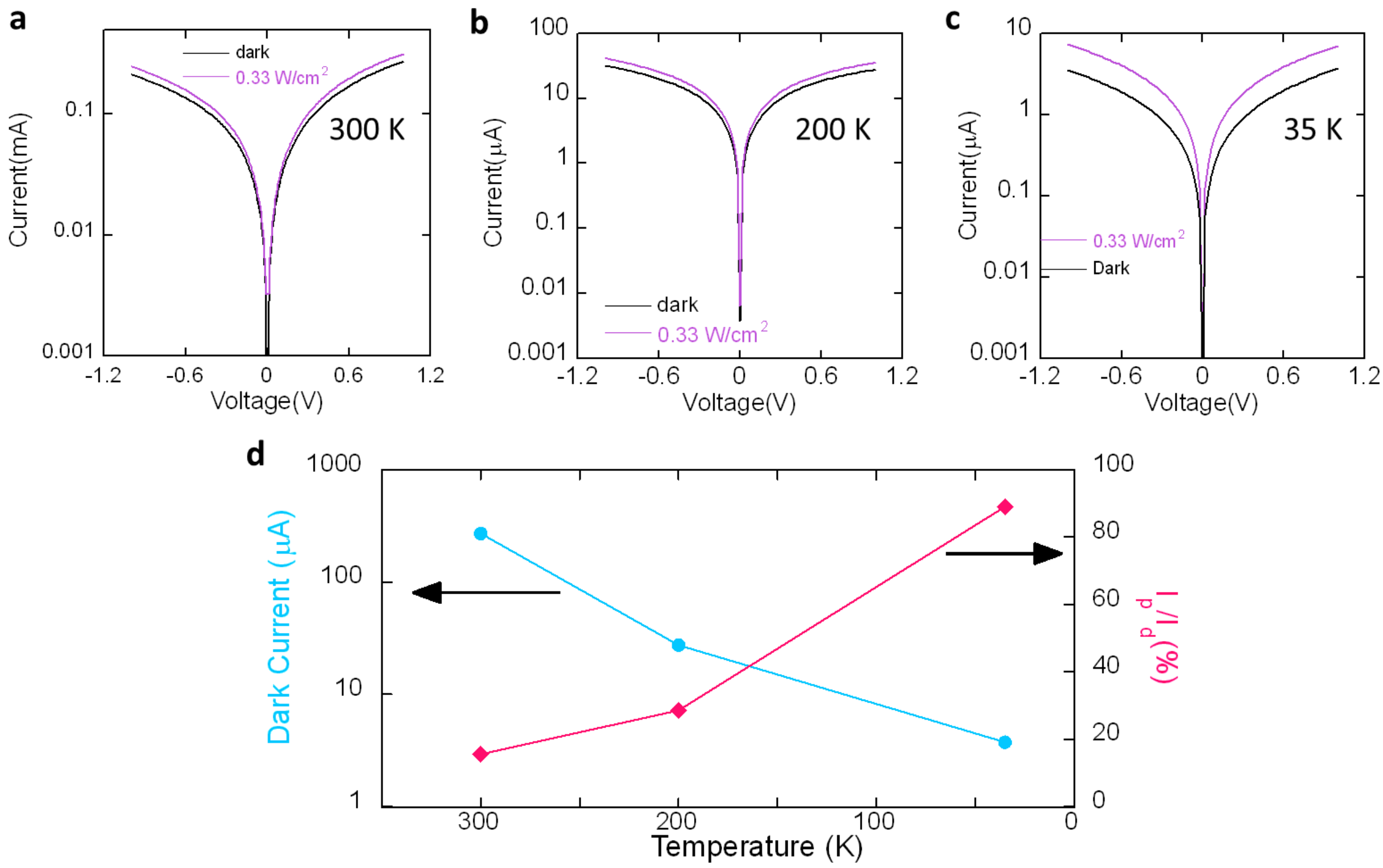

**Figure S16**: The temperature dependence of photo response. **(a-c)** I-V characteristics of a typical horizontal device measured at 300 K, 200 K and 35 K respectively. **(d)** The dark current and the percentage of photocurrent to dark current ratio ($I_p/I_d = I_{light} - I_{dark}/I_{dark}$) is clearly seen to be increasing as the temperature decreases. Here the dark current and ratio of $I_p/I_d$ is calculated at 1V of bias.

## 10. Thermal Imaging Set-up:

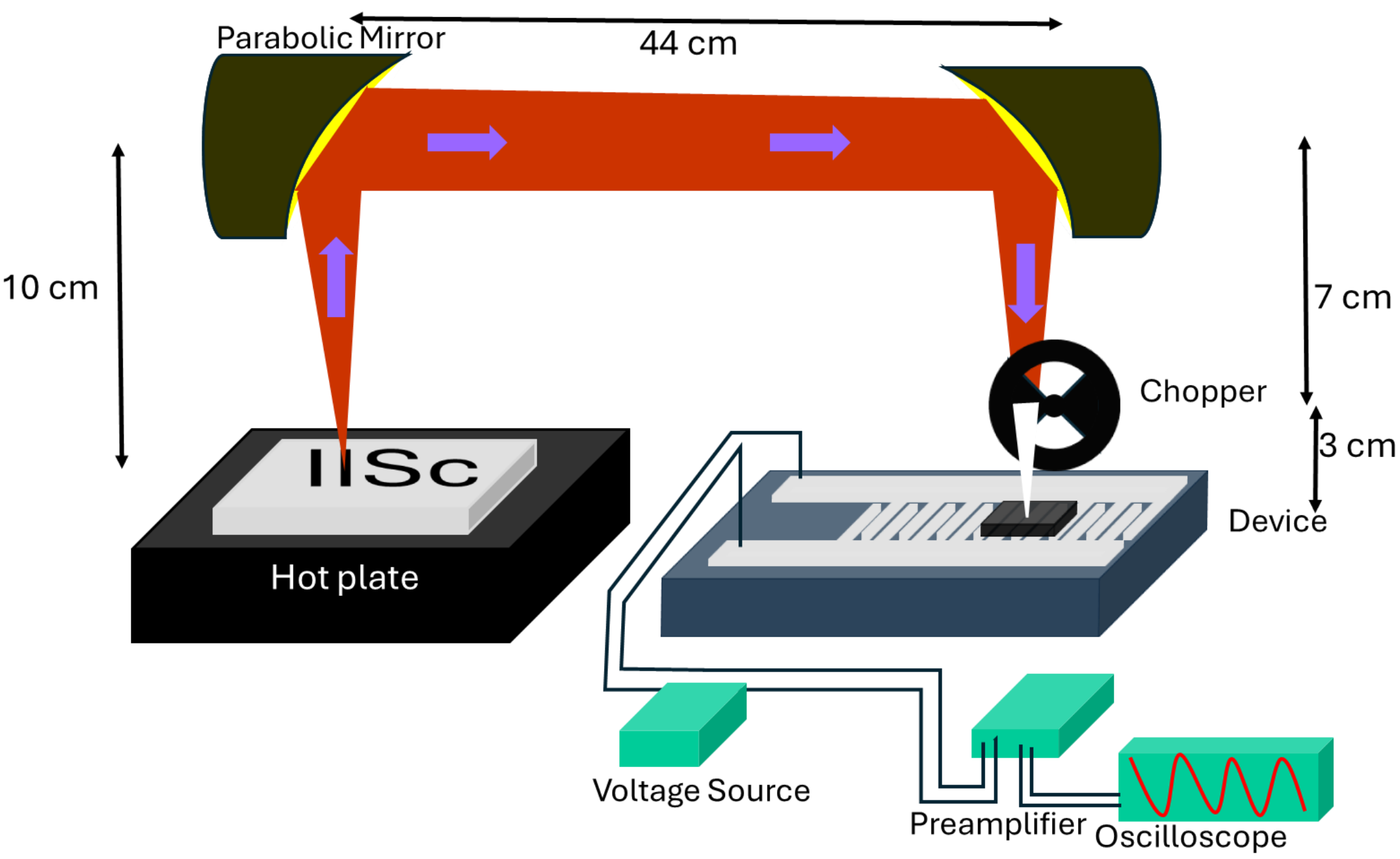

**Figure S17:** Imaging using Horizontal photodetector: **(a)** The set up for imaging of is made with a hot plate covered with aluminum foil and the foil was cut with the letters desired to be imaged. The ceramics was open to emission to the parabolic mirror in the place of the letters. And the remaining area was covered with aluminum foil indicated by white layer on the hot plate. The parabolic mirrors have a focal distance of 10 cm. That is why the objects are the device is kept at a distance of 10 cm from the parabolic mirror. The relative dimensions of the object drawn here are not up to the scale.

Since the optical band edge of the Se-$I_2$ material lies deep in the infrared, we attempted thermal imaging using a commercial infrared emitter. For this purpose, the ceramic element of an "ANTECH" hot plate was employed as a broadband infrared source. The surface of the hot plate was covered with aluminium foil, which was patterned by cutting out the desired letters or objects to be imaged. This approach ensured that only specific regions of the ceramic were exposed to radiative emission, while the rest of the hot plate surface was effectively blocked by the aluminium mask. In this way, the ceramic acted as a spatially patterned infrared emitter, with the exposed regions functioning as the object to be imaged.

The emitted thermal radiation was collected by a parabolic mirror positioned at its focal distance of 10 cm from the hot plate (Figure 4.13). The first parabolic mirror collimated the divergent emission, transforming it into a parallel beam. A second parabolic mirror, also with a 10 cm focal distance, was then used to refocus this collimated radiation into a tight spot on the photodetector surface.

To enhance signal-to-noise ratio, the infrared emission was modulated using an optical chopper operating at a low frequency of 4 Hz. A low chopping frequency was deliberately chosen to maximize the signal collected from the broadband thermal source and to ensure compatibility with the detector response time. The modulated signal was subsequently fed into a current preamplifier, which boosted the weak photocurrent signal without significantly distorting it. The amplified signal was recorded on a digital oscilloscope. Simultaneously, the oscilloscope data was synchronized with the (x, y) coordinates of a motion controller, which was used to raster-scan the detector across the image plane. By mapping the signal intensity as a function of spatial coordinates, the thermal image of the patterned letters could be reconstructed.

To minimize thermal interference, the two parabolic mirrors were placed at a safe working distance of ~44 cm from each other. This ensured that the photodetector and associated electronics were not directly affected by the heat from the hot plate. The purple arrows in Figure 4.13 schematically indicate the path of the infrared radiation through the optical system. Overall, this setup demonstrates the feasibility of infrared thermal imaging using Se-$I_2$ photodetectors.

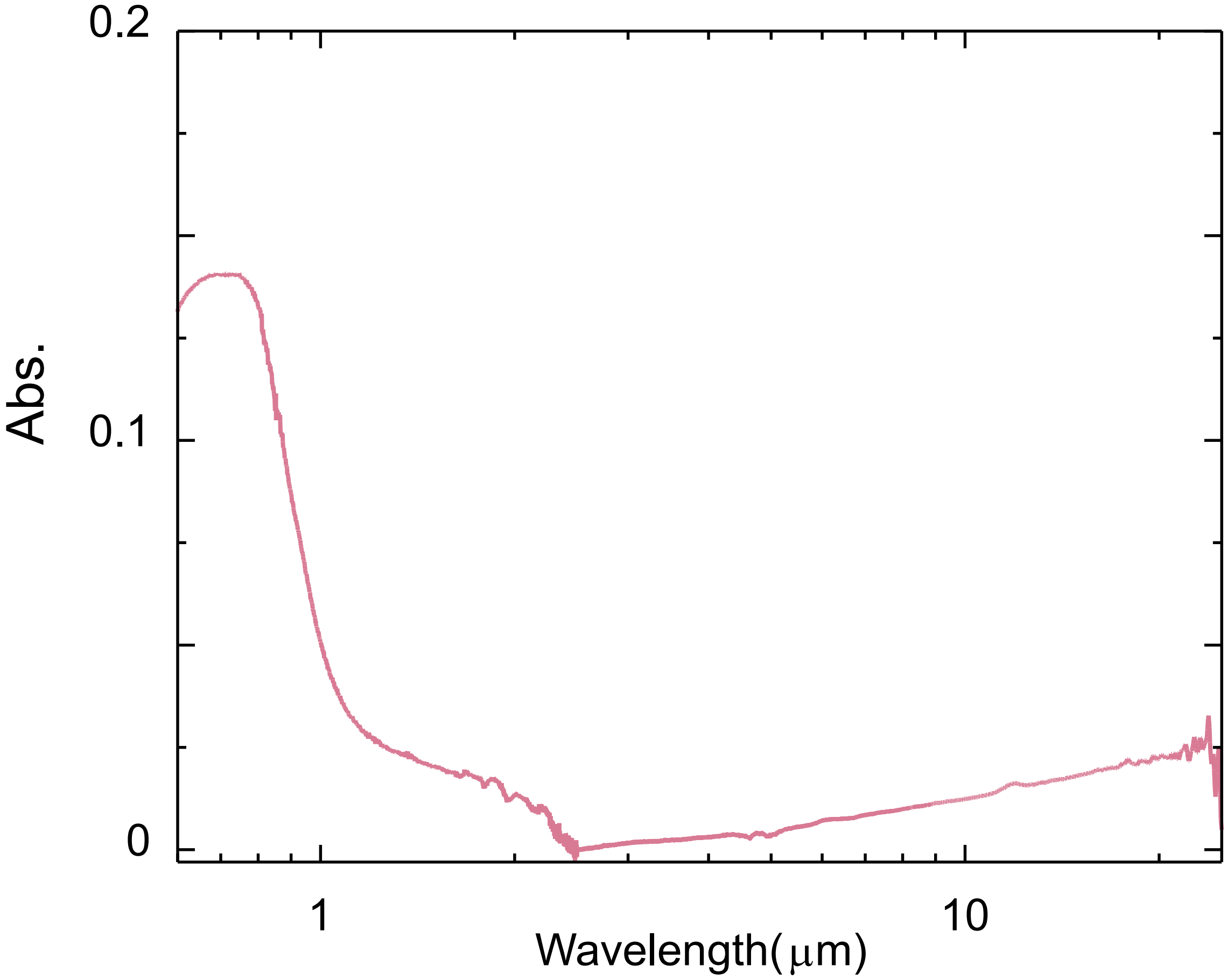

**Figure S18:** Absorbance spectra of Se-$I_2$ covering both visible and infrared region. The spectra from 0.4 $\mu$m to 25 $\mu$m is taken with Perkin Elmer Lamda 750 instrument in transmission mode with thin film of Se-$I_2$. Similarly the spectra from 2.5 $\mu$m to 25 $\mu$m is collected with a Fourier Transform-infrared spectrometer|(FT-IR) in transmission mode.

## 11. Experimental:

**11.1. Synthesis and Basic Characterizations**: The material has been made by adding 0.8 mmol (200 mg) iodine ($I_2$) to 0.8 mmol (64 mg) of Selenium (Se) powder in a round bottom flask in inert condition (Argon in Schlenk line), gently heating it up to 70ºC to liquify the mixture, after complete liquification, the mixture has been poured into respective substrate for characterization, For XPS and SEM, the samples have been poured over a 1 cm × 1cm silicon wafer. For recording XRD pattern, and Raman shifts the samples have been drop cast on a glass slide.

For the TGA experiment the samples have been put in a sealed ampule made of aluminum. 5 mg samples have been enclosed in a aluminum ampule, to ensure leakage free experiment since the material is prone to evaporate iodine upon heating.

**11.2. Horizontal Device Fabrication:** 32 mg of Selenium and 100 mg of iodine is weighed and heated into 70°C with glass vial, A glass coated platinum interdigit has been prepared with covering with a capton tape, leaving a area of 1.5 mm×1.5 mm area where the sample must be poured for making the device. The hot melt SeI2 was poured over the open area of the interdigit. A glass slide was pulled over the surface of the electrode to make a thin film. This shows metallic luster on its surface. The device was kept inside a closed cylindrical container of base dimeter 2.2 cm and height of 2.1 cm for 2 weeks. After 2 weeks a bit 25 mg of iodine was weighed and put into container and closed it for iodine infusion into the device to maintain a equilibrium concentration of iodine in the $Se_nI_2$ sample. After 4 weeks of iodine infusion the device was taken out in the air. There was no iodine leakage from the device after keeping it outside in the air for 6 hours. The device was monitored for 12 months after preparation. There was no weight loss of the device after 2 months of keeping it in the open air, noticed by weighing the device with a weight balance machine.

**11.3. XRD:** Film of $Se_nI_2$ sample was prepared on a glass slide. The XRD pattern was recorded using a Malvern Panalytical Empyrean diffractometer.

**11.4. TEM:** A thin flake from the sample was stripped with a scotch tape, then transferred to the TEM grid. A TITAN machine was used for transmission electron microscopy experiments

**11.5. SEM:** The liquid sample was drop cast on a silicon wafer and allowed to solidify. An ultra55 Zeiss instrument was used for all scanning electron microscopic experiments

**11.6. Optoelectronic Measurements:** A Keithley 2450 source meter was used for probing the electronics from all two terminal detector devices. Lasers of different wavelength (405 nm, 532 nm, 630 nm, 808 nm, 1550 nm) has been used with neutral density filters to control the intensity. A silver mirror is used to guide the laser beams to the device active area. These experiments has been done a vibration free optical table to ensure noise removal. A optical chopper 2/2 wheel with a chopper controller was used to modulate the light chopping frequency. For milliseconds of response time evaluation, a current preamplifier with a Mokugo lock-in amplifier was used to record the photo response while chopping the laser at the desired frequency.